\documentclass[twocolumn,showpacs,prb,amsfonts,amsmath,amssymb,floatfix]{revtex4-2} 
\usepackage[dvipdfmx]{graphicx}
\usepackage{bmpsize}
\usepackage{color}
\usepackage{hhline}
\usepackage{mathrsfs}
\usepackage{dcolumn}
\usepackage{bm}
\usepackage{multirow}
\usepackage{booktabs}
\usepackage{afterpage}
\usepackage{amsmath}
\usepackage{braket}
\usepackage{subfigure}
\usepackage{newtxtext} 
\usepackage{txfonts} 

\begin{document}

\title{Development of an efficient impurity solver in dynamical mean field theory for multi-band systems: \\The iterative perturbation theory combined with the parquet equations}

\author{Ryota Mizuno}\email{mizuno@presto.phys.sci.osaka-u.ac.jp}
\author{Masayuki Ochi}
\author{Kazuhiko Kuroki}

\affiliation{ Department of Physics, Osaka University, 1-1 Machikaneyama, Toyonaka, Osaka 560-0043, Japan}

\date{\today}
\begin{abstract}
  Although several impurity solvers in the dynamical mean field theory (DMFT) have been proposed, 
  especially in multi-band systems, 
  there are practical difficulties arising from a trade-off between numerical costs and reliability. 
  In this study, we re-interpret the iterative perturbation theory (IPT) as an approximation which captures the strong correlation effects by mimicking the particular frequency structures of the exact full vertex,
  and extend it such that it can have efficiency and reliability simultaneously by modifying IPT vertex using the parquet equations. 
  We apply this method to several models to evaluate their validity.
  We confirm that our method shows good agreements with the numerically exact continuous-time quantum Monte Carlo method in the single-site DMFT calculation. 
\end{abstract}
\maketitle

\section{introduction}
Strongly correlated systems exhibit many exciting phenomena such as high-temperature superconductivity, 
metal-insulator transition and so on.
However, these mechanisms cannot be understood in detail.
These phenomena emerge in the region 
where 
both the perturbation expansions from  the itinerant and localized pictures break down.
In addition to this non-perturbative nature, 
we need to consider multi-orbital or multi-site degrees of freedom.
Due to these complexities,
these phenomena are still unresolved problems 
even though several decades have passed since their discovery.
It is one of the central issues in condensed matter physics to understand the strong correlation effects.

Dynamical mean field theory~(DMFT)~\cite{RevModPhys.68.13} is one of the most powerful methods to study the strongly correlated systems. 
DMFT is a method in which the lattice problem is solved non-perturbatively by mapping it onto an impurity problem.
DMFT can treat  
the temporal fluctuation correctly 
and 
can connect the itinerant and localized limits smoothly.
Although DMFT has these excellent features, 
it cannot describe the phenomena such as anisotropic superconductivity or pseudo gap since the spatial fluctuation is ignored.
To resolve this problem, 
extensions which take into account the spatial fluctuation in DMFT were developed~\cite{RevModPhys.77.1027,RevModPhys.90.025003}.  
Further, formalisms to combine DMFT with {\it ab initio} methods were established, and so DMFT is nowadays applied to various realistic calculations.

As mentioned above,
in DMFT,
the lattice problem is solved by mapping it onto an impurity problem. 
The most widely used impurity solver is the continuous-time quantum Monte Carlo method~(CT-QMC)~\cite{PhysRevLett.97.076405,PhysRevB.76.235123,PhysRevB.74.155107,doi:10.1143/JPSJ.76.114707,PhysRevB.72.035122,Rubtsov2004,PhysRevB.89.195146},
which is numerically exact.
However, 
in multi-band systems,
it can suffer from a serious sign problem
and the numerical cost necessary to obtain results with sufficiently small statistical errors grows rapidly with increasing the number of bands.
In addition,
the numerical cost can also increase due to 
the exponential growth of the Fock space of the multi-band impurity problem in the hybridization expansion of CT-QMC~(CT-HYB)~\cite{PhysRevLett.97.076405,PhysRevB.76.235123,PhysRevB.74.155107}
and 
the growth of the average perturbation order in the interaction expansion~(CT-INT)~\cite{PhysRevB.72.035122,Rubtsov2004,PhysRevB.89.195146}.
Another exact impurity solver is the exact diagonalization method~(ED)~\cite{PhysRevLett.72.1545,PhysRevB.86.165128,PhysRevB.81.054513,PhysRevB.84.180505,Liebsch_2011}.
Although the formalism of ED itself is exact, we need to discretize the electron bath in actual calculations.
When we apply it to multi-band systems, the discretization error can become worse because of the trade-off relation between the numbers of the impurity orbitals and bath levels. 
Since at least two bath levels per impurity orbital are required to obtain reliable results~\cite{Liebsch_2011,PhysRevB.81.054513,PhysRevB.84.180505}, 
the numerical cost grows exponentially with increasing the number of bands.
Moreover, the broadening procedures to obtain the continuous spectrum from the resulting discrete spectrum have ambiguity.
To avoid this practical difficulty, 
it is often necessary to use 
a numerically low-cost 
approximation method as an impurity solver.
The iterative perturbation theory (IPT)~\cite{doi:10.1143/PTPS.46.244, doi:10.1143/PTP.53.970, doi:10.1143/PTP.53.1286, Yamada4, PhysRevB.45.6479},
which we bring up in this paper,
is one of these methods.

The original IPT was developed as a very simple approximation,
in which the self energy is calculated by the second-order perturbation. 
This self energy coincidently reproduces the atomic (strong correlation) limit in the electron-hole symmetric case.
Hence, in this condition,
IPT is a highly useful method which can connect the weakly and strongly correlated regime
even though it is a perturbation scheme.
Later, the modified-IPT, which is an extended version for an electron-hole asymmetric case, was developed~\cite{PhysRevLett.77.131,PhysRevB.55.16132,PhysRevB.86.085133}.
In this method, the self energy is parameterized so that it reproduces the exact solutions in the high-frequency and the atomic limits.
By this improvement,
IPT became able to be applied to the electron-hole asymmetric systems.
Further improvement for multi-orbital systems has been made~\cite{Saso_2001,doi:10.1143/JPSJ.72.777,PhysRevLett.91.156402,Dasari2016}.
Similarly to the modified-IPT,
the parameters are determined such that the self energy reproduces the high-frequency limit.
In multi-orbital systems, however,
since the exact solution in the atomic limit cannot be obtained in a simple form,
the self energy is determined such that it reproduces the approximate solution in the atomic limit.
Hence the scopes of the application of these methods are quite restricted.

As described above, IPT has been regarded as a method which interpolates the weak and strong correlation limits. 
In this study, 
we provide IPT with a new interpretation 
in which IPT captures the strong correlation effects by mimicking the particular frequency structures of the exact full vertex,
and extend the method such that it can be applied to multi-band systems. 
We validate this method by applying it to several models and comparing with the numerically exact CT-QMC method.


This paper is organized as follows.
In Sec.~\ref{sec:2020-12-25-03-56},
we define the models and outline the Green's function method.
We describe in  Sec.~\ref{sec:2020-12-25-04-04} the novel method developed in the present study.
Results are shown in Sec.~\ref{sec:2020-12-25-04-05}.
The discussion is presented in Sec.~\ref{sec:2021-04-07-14-03}.
The conclusion is given in Sec.~\ref{sec:2020-12-25-04--13}.

\section{Model and Green's function}\label{sec:2020-12-25-03-56}
\subsection{Definitions}
We consider the Hubbard model for multi-band systems described by the following Hamiltonian.
\begin{align}
  H 
  =&
  \sum_{ij}\sum_{\alpha\beta}t_{ij,\alpha\beta}c^{\dagger}_{i\alpha}c_{j\beta} 
  +
  \dfrac{1}{4} \sum_{i} \sum_{\alpha\beta\gamma\lambda} U_{\alpha\beta\gamma\lambda} c^{\dagger}_{i\alpha}c^{\dagger}_{i\lambda}c_{i\gamma}c_{i\beta}, 
  \label{eq:2020-05-12-23-57}
\end{align}
where 
the subscripts with Roman letters indicate unit cells 
and
Greek letters the set of the degree of freedom of spin, orbital, and site.
$t_{ij,\alpha\beta}$ is the hopping integral 
and 
$U_{\alpha\beta\gamma\lambda}$ is the Coulomb repulsion.
$c_{i\alpha}^{(\dagger)}$ is the annihilation (creation) operator.

The $n$-particle Green's function is defined as 
\begin{align}
  G^{(n)}_{i_{1},\cdots,i_{2n},\alpha_{1},\cdots,\alpha_{2n}}&(\tau_{1},\cdots,\tau_{2n}) \nonumber \\
  &\hspace{-50pt}=
  (-1)^{n} \bigl<T[
    c_{i_{1}\alpha_{1}}(\tau_{1})c^{\dagger}_{i_{2}\alpha_{2}}(\tau_{2})
    \cdots 
  c_{i_{2n-1}\alpha_{2n-1}}(\tau_{2n-1})c^{\dagger}_{i_{2n}\alpha_{2n}}(\tau_{2n})]
  \bigr>,
  \label{eq:2020-10-08-14-25}
\end{align}
where
$c^{(\dagger)}(\tau)=e^{\tau H}c^{(\dagger)}e^{-\tau H}$ is the Heisenberg representation of creation (annihilation) operators.
$\braket{A}={\rm Tr}(e^{-\beta H}A)/Z$ is the statistical average of $A$
and 
$Z={\rm Tr}(e^{-\beta H})$ is the partition function.

In the presence of the time and lattice translational invariance, 
one-particle Green's function~[$n=1$ in Eq.~(\ref{eq:2020-10-08-14-25})] in the momentum space can be written as 
\begin{align}
  G_{\alpha\beta}(\bm{k},\tau)
  \equiv
  G^{(1)}_{\alpha\beta}(\bm{k},\tau)
  =&
  -\bigl<T c_{\bm{k}\alpha}(\tau)c^{\dagger}_{\bm{k}\beta} \bigr>
\end{align}
where $\bm{k}$ denotes the momentum.
The Fourier transformation in terms of the imaginary time is expressed as 
\begin{align}
  G_{\alpha\beta}(\bm{k},\tau) =& \dfrac{1}{\beta}\sum_{n} G_{\alpha\beta}(\bm{k},i\omega_{n}) e^{-i\omega_{n}\tau}
  \label{eq:2020-10-08-14-50} \\
  G_{\alpha\beta}(\bm{k},i\omega_{n}) =& \int d\tau  G_{\alpha\beta}(\bm{k},\tau)e^{i\omega_{n}\tau}
  \label{eq:2020-10-08-14-51}
\end{align}
where 
$\omega_{n}=(2n+1)\pi T$ with $n \in {\mathbb Z}$ is a fermionic Matsubara frequency
[$\nu_{m}=2m \pi T$ introduced later is a bosonic Matsubara frequency]. 
$G(\bm{k},i\omega_{n})$ can be derived in the following form. 
\begin{align}
  \hat{G}(k) 
  =&
  \bigl[ (i\omega_{n} + \mu)\hat{I} - \hat{\epsilon}_{\bm{k}} - \hat{\Sigma}(k) \bigr]^{-1},
  \label{eq:2020-10-08-14-03}
\end{align}
where 
$\mu$ is the chemical potential and 
$k=(\bm{k},i\omega_{n})$ is the generalized fermionic momentum
[$q=(\bm{q},i\nu_{m})$ introduced later denotes the generalized bosonic momentum].
$\hat{\epsilon}_{\bm{k}}=N_{\bm{k}}^{-2}\sum_{ij}\hat{t}_{ij}e^{i(\bm{R}_{i}-\bm{R}_{j})\cdot\bm{k}}$ is the band dispersion
and 
$\hat{\Sigma}(k)$ is the self energy.
These quantities are matrices in terms of the band index
and $\hat{I}$ is the unit matrix.

Similarly to the one-particle case,
in the presence of the time and lattice translational invariance, 
the two-particle Green's function~[$n=2$ in Eq.~(\ref{eq:2020-10-08-14-25})] in the momentum space can be written as 
\begin{align}
  G^{(2)}_{\alpha\beta\gamma\lambda}(\bm{k},\bm{k}',\bm{q}, \tau_{1},\tau_{2},\tau_{3}) 
  =&
  \Bigl< T c_{\bm{k}\alpha}(\tau_{1})c^{\dagger}_{\bm{k}+\bm{q}\beta}(\tau_{2})c_{\bm{k}'+\bm{q}\lambda}(\tau_{3})c^{\dagger}_{\bm{k}'\gamma} \Bigr> .
  \label{eq:2020-10-08-23-02}
\end{align}
Fourier transformation is given by 
\begin{align}
  \hat{G}^{(2)}&(\bm{k},\bm{k}',\bm{q}, \tau_{1},\tau_{2},\tau_{3}) \nonumber \\
  =&
  \dfrac{1}{\beta^{3}} \sum_{nn'm} \hat{G}^{(2)}(\bm{k},\bm{k}',\bm{q}, i\omega_{n},i\omega_{n'},i\nu_{m})
  e^{-i\omega_{n} \tau_{1}} e^{i(\omega_{n}+\nu_{m})\tau_{2}} e^{-i(\omega_{n'}+\nu_{m})\tau_{3}} .
  \label{eq:2020-10-08-23-03}
\end{align}
The two-particle Green's function can be divided into two parts:
disconnected and connected terms.
\begin{align}
  &G^{(2)}_{\alpha\beta\gamma\lambda}(k,k',q) \nonumber \\
  &=
  G_{\alpha\beta}(k) G_{\lambda\gamma}(k') \delta_{q,0} 
  -
  G_{\alpha\gamma}(k) G_{\lambda\beta}(k+q) \delta_{kk'}
  \nonumber \\
  &+
  \hspace{-5pt}\sum_{\alpha'\beta'\gamma'\lambda'}\hspace{-5pt}
  G_{\alpha\gamma'}(k) G_{\lambda'\beta}(k+q) F_{\gamma'\lambda'\alpha'\beta'}(k,k',q) G_{\alpha'\gamma}(k') G_{\lambda\beta'}(k'+q),
  \label{eq:2020-10-08-23-04}
\end{align}
where  
$\hat{F}$ is called the full vertex.
Introducing the irreducible susceptibility 
\begin{align}
  \chi_{0,\alpha\beta\gamma\lambda} (k,k',q)
  =&
  -G_{\alpha\gamma}(k) G_{\lambda\beta}(k+q) \delta_{kk'}, 
  \label{eq:2020-10-08-23-05}
\end{align}
we can define the generalized susceptibility as 
\begin{align}
  \hat{\chi}_{\rm G}(k,k',q)
  =&
  \hat{\chi}_{0}(k,k',q) - \hat{\chi}_{0}(k,q)\hat{F}(k,k',q)\hat{\chi}_{0}(k',q) .
  \label{eq:2020-10-08-23-06}
\end{align}
Also we can write the self energy by using full vertex as 
\begin{align}
  \Sigma_{\alpha\beta}(k) 
  =&
  \Sigma_{\alpha\beta}^{\rm HF} 
  + 
  \dfrac{1}{2}\sum_{\gamma\lambda}\sum_{k',q} [\hat{F}(k,k',q) \hat{\chi}_{0}(k',q) \hat{U}]_{\alpha\gamma\beta\lambda} G_{\gamma\lambda}(k+q) ,
  \label{eq:2020-10-08-23-07}
\end{align}
where 
$\Sigma^{\rm HF}$ is the Hartree-Fock term.

\subsection{Parquet formalism}
\begin{figure*}[t]
  \centering
  {\includegraphics[width=140mm,clip]{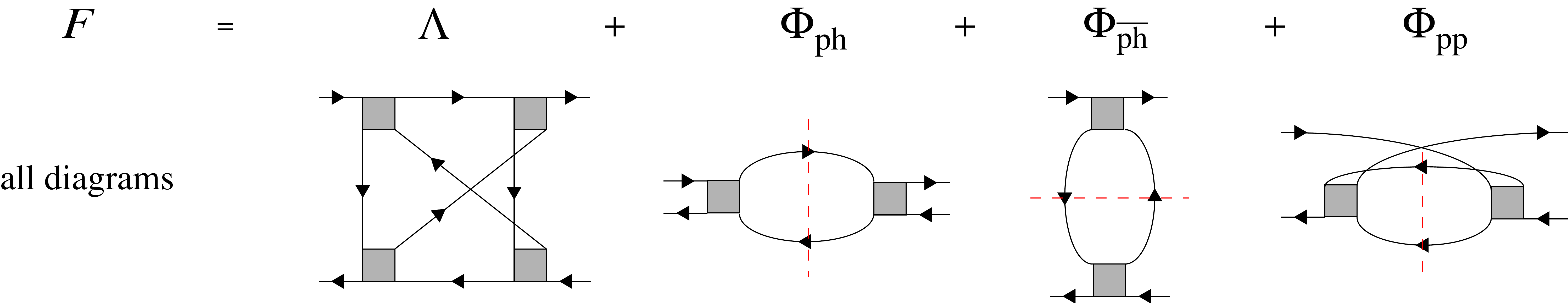}} 
  \caption{The decomposition of the full vertex.
    The full vertex can be divided into four parts:
    the fully irreducible part ($\Lambda$) and the reducible parts ($\Phi_{l}$,  $l=$ ph, ${\rm \overline{ph}}$ pp).  
  }
  \label{fig:2020-05-09-22-21}
\end{figure*}

To consider the diagrammatic structure of the full vertex $F$, 
we have to define the irreducible susceptibilities concerning the following three channels (${\rm ph, \overline{ph}, pp }$).
\begin{align}
  \chi_{0,\alpha\beta\gamma\lambda}(k,k',q) =& 
  \begin{cases}
    - G_{\alpha\gamma}(k)G_{\lambda\beta}(k+q) \delta_{kk'} \hspace{5pt} &(\text{ph}) \\
    G_{\alpha\beta}(k) G_{\lambda\gamma}(k') \delta_{q0}    &({\overline{\rm ph}}) \\
    G_{\alpha\gamma}(k)G_{\beta\lambda}(-k-q)\delta_{kk'}     &(\text{pp})
  \end{cases}.
  \label{eq:2020-10-08-23-42} 
\end{align}
The ph channel in Eq.~(\ref{eq:2020-10-08-23-42}) is the same as Eq.~(\ref{eq:2020-10-08-23-05}).
The full vertex $F$ can be divided into four parts, 
\begin{align}
  \hat{F} &= \hat{\Lambda} + \hat{\Phi}_{\rm ph} + \hat{\Phi}_{\rm \overline{ph}} + \hat{\Phi}_{\rm pp},
  \label{eq:2020-05-09-22-20}
\end{align}
where 
$\hat{\Phi}_{l}$ $(l={\rm ph,\overline{ph},pp})$
is the set of reducible diagrams in channel $l$,
and 
$\hat{\Lambda}$ is the set of fully irreducible diagrams. 
The diagrammatic representation is shown in Fig.~\ref{fig:2020-05-09-22-21}.
Since there is no diagram which simultaneously satisfies reducibility  in two or more channels, 
we can write 
\begin{align}
  \hat{F} =& \hat{\Gamma}_{l} + \hat{\Phi}_{l} 
  \label{eq:2020-05-09-22-53} \\
  \hat{\Gamma}_{l} =& \hat{\Lambda} + \hat{\Phi}_{l_{1}} + \hat{\Phi}_{l_{2}} \hspace{10pt} (l\neq l_{1} \neq l_{2}) 
  \label{eq:2020-05-09-22-54} \\
  \hat{\Phi}_{l} =& 
  -\hat{\Gamma}_{l}\hat{\chi}_{0} \hat{F} = -\hat{\Gamma}_{l}\hat{\chi}_{l} \hat{\Gamma}_{l}, 
\end{align}
where 
$\hat{\Gamma}_{l}$ is the set of diagrams irreducible in channel $l$ 
and 
is called the irreducible vertex in $l$.
$\hat{\chi}_{l}$ are given by
\begin{align}
  \hat{\chi}_{l} =& \hat{\chi}_{0} - \hat{\chi}_{0}\hat{\Gamma}_{l}\hat{\chi}_{l} = \hat{\chi}_{0} - \hat{\chi}_{0}\hat{F}\hat{\chi}_{0} . \label{eq:2020-05-12-14-40} 
\end{align}
From 
Eqs.~(\ref{eq:2020-05-09-22-53})$-$(\ref{eq:2020-05-12-14-40}), which are called the parquet equations~\cite{e_023_03_0489,PhysRevB.86.125114,Janis_1998,PhysRevB.60.11345}, 
we can calculate $\hat{F}$ exactly if we know the exact $\hat{\Lambda}$.
However, it is very difficult to obtain the exact $\hat{\Lambda}$ 
and the procedure to obtain $\hat{\Phi}_{l}$ is numerically very expensive. 
Thus, 
some approximations or simplifications have been proposed 
\cite{doi:10.1143/JPSJ.79.094707,PhysRevB.75.165108,PhysRevB.83.035114}
[see Appendix~\ref{sec:2020-10-09-00-10}].

\section{Novel impurity solver: IPT+parquet}\label{sec:2020-12-25-04-04}

In this section,
we develop a novel impurity solver by re-interpreting and extending IPT.
First, in Sec.~\ref{sec:2020-11-06-17-44}, 
we introduce our extension of IPT. 
After that in Sec.~\ref{sec:2021-04-22-01-15}, 
we explain our new interpretation of IPT, which is the basis of the extension.
Other theoretical details are in Sec.~\ref{sec:2021-06-24-14-06}.

\subsection{Extension of IPT}\label{sec:2020-11-06-17-44}

In IPT, 
the correlation part of the self energy is approximated as 
\begin{align}
  \hat{\Sigma}_{\rm IPT}(\omega_{n})
  =&
  [\hat{I}-\hat{B}\hat{\Sigma}^{\rm 2nd}(i\omega_{n})]^{-1}\hat{A} \hat{\Sigma}^{\rm 2nd} (i\omega_{n}) \label{eq:2020-06-02-20-56} 
\end{align}
\begin{align} 
  {\Sigma}^{\rm 2nd}_{\alpha\beta} =&
  T^{2}\sum_{\gamma\lambda}\sum_{\omega_{n'}\nu_{m}}[\hat{U}\hat{\chi}_{0}(\omega_{n'},\nu_{m})\hat{U}]_{\alpha\gamma\beta\lambda}G_{0,\gamma\lambda}(\omega_{n}+\nu_{m})
  \label{eq:2020-06-03-19-51}\\
  &\chi_{0,\alpha\beta\gamma\lambda}(\omega_{n},\nu_{m}) = -G_{0,\alpha\gamma}(\omega_{n})G_{0,\lambda\beta}(\omega_{n}+\nu_{m})
  \label{eq:2021-06-29-00-13}\\
  &\hat{G}_{0}(i\omega_{n}) = [ (i\omega_{n} + \mu_{0})\hat{I} - \hat{\Delta}(i\omega_{n}) - \hat{\Sigma}^{\rm HF} ]^{-1},
  \label{eq:2021-04-02-23-00}
\end{align}
where $\mu_{0}$, $\hat{\Delta}(i\omega)$, and $\hat{\Sigma}^{\rm HF}$ are the pseudo chemical potential, the hybridization function, and the Hartree-Fock term in the self energy, respectively.
The parameters $\hat{A},\hat{B}$ 
are determined such that one reproduces the exact solutions in the high frequency and atomic limits: 
\begin{align}
  \hat{A} =& \dfrac{n(1-n)}{n_{0}(1-n_{0})}\hat{I}, \hspace{10pt} \hat{B} = \dfrac{(1-2n)U + \mu_{0} - \mu}{n_{0}(1-n_{0})U^{2}}\hat{I},
  \label{eq:2021-04-02-22-59} 
\end{align}
where 
$n_{0}$ and $n$ are the electron numbers evaluated from $\hat{G}_{0}(i\omega_{n})$ and $\hat{G}(i\omega_{n})$, respectively.
Although this is the IPT formalism for the single-band systems, 
we intentionally write it in the matrix form for the extension below.

We extend the IPT as follows.
\begin{align}
  \hat{\Sigma}^{\rm CR}_{\rm IPT+parquet}(\omega_{n}) =& [\hat{I} - \hat{B}\hat{\Sigma}^{\rm CR}_{0}(\omega_{n})]^{-1}\hat{A} \hat{\Sigma}^{\rm CR}_{0}(\omega_{n}) 
  \label{eq:2020-08-28-02-01} 
\end{align}
\begin{align}
  {\Sigma}^{\rm CR}_{0,\alpha\beta}(\omega_{n})
  =&
  T^{2}\sum_{\gamma\lambda}\sum_{\omega_{n}'\nu_{m}}
  [\hat{F}_{0}(\omega_{n},\omega_{n'},\nu_{m}) 
    \hat{\chi}_{0}(\omega_{n'},\nu_{m})
  \hat{U} ]_{\alpha\gamma\beta\lambda} \nonumber \\
  &\hspace{30pt}\times G_{0,\gamma\lambda}(\omega_{n}+\nu_{m})
  \label{eq:2020-06-04-14-05} \\
  &\chi_{0,\alpha\beta\gamma\lambda}(\omega_{n},\nu_{m}) = -G_{0,\alpha\gamma}(\omega_{n})G_{0,\lambda\beta}(\omega_{n}+\nu_{m})
  \label{eq:2021-06-29-00-25}\\
  &\hat{G}_{0}(\omega_{n}) = [i\omega_{n}\hat{I}+\hat{\mu}_{0}- \hat{\Delta}(\omega_{n}) - \hat{\Sigma}^{\rm HF} ]^{-1}
  \label{eq:2020-07-02-19-01} 
\end{align} 
\begin{align}
  \hat{F}_{0}&(\omega_{n},\omega_{n'},\nu_{m}) 
  \nonumber \\
  =&
  \hat{U} + \hat{\Phi}_{\rm ph}(\nu_{m}) + \hat{\Phi}_{\rm \overline{ph}}(\omega_{n}-\omega_{n'}) + \hat{\Phi}_{\rm pp}(\omega_{n}+\omega_{n'}+\nu_{m}),
  \label{eq:2020-08-28-01-28}
\end{align} 
where $\hat{F}_{0}$ is an approximate full vertex.
To obtain $\hat{F}_{0}$, we employ the simplified parquet method developed in Ref.~\cite{doi:10.1143/JPSJ.79.094707}~[We explain the reason why we employ the simplified parquet method in Sec.~\ref{sec:2021-04-22-01-15}]. 
Hence, we call this ``IPT+parquet method''.
Since the simplified parquet method in Ref.~\cite{doi:10.1143/JPSJ.79.094707} supports only the single-band calculations,
we extend it for the multi-band calculations and its detailed procedure to obtain $\hat{F}_{0}$ is shown in Appendix~\ref{sec:2020-10-09-00-10}.
In practical calculation, however,
we omit the contribution from pp channel $\hat{\Phi}_{\rm pp}$ when calculating the self energy in Eq.~(\ref{eq:2020-07-02-19-01}) since $\hat{\Phi}_{\rm pp}$ tends to be overestimated.
We add a band index to the pseudo chemical potential $\mu_{0}$ in Eq.~(\ref{eq:2021-04-02-23-00}), 
and so $\hat{\mu}_{0}$ in Eq.~(\ref{eq:2020-07-02-19-01}) is a diagonal matrix.
The reason for this modification and the conditions for the parameters $\hat{A},\hat{B}$, and $\hat{\mu}_{0}$ are discussed later in Sec.~\ref{sec:2021-06-24-14-06}.

This extension is based on the interpretation in which IPT captures the strong correlation effects by mimicking the particular frequency structures of the exact full vertex.
We explain this new interpretation in detail in the next section.

\subsection{Re-interpretation of IPT}\label{sec:2021-04-22-01-15}
\begin{figure}[]
  \centering
  {\includegraphics[width=50mm,clip]{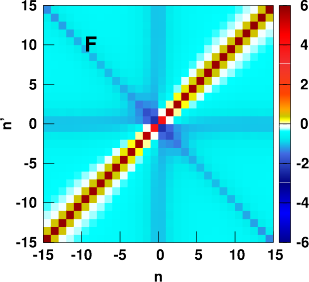}} 
  \caption{The frequency dependence of the full vertex in the charge channel obtained in QMC as an impurity solver.
    The bare interaction $U$ is subtracted.
    The calculations have been performed for the Hubbard model on a square lattice with nearest-neighbor hopping $t$ at $T/t=0.4$, $U/t=5.08$.
    The intensity is given in unit of $4t$.
    This figure is taken from Ref.
    \cite{RevModPhys.90.025003}.
  } 
  \label{fig:2020-06-14-00-39}
\end{figure}

To present our new interpretation of IPT,
we explain the frequency structure of the exact full vertex~\cite{RevModPhys.90.025003,PhysRevB.94.235108,PhysRevB.86.125114}.
Figure~\ref{fig:2020-06-14-00-39} shows the full vertex in the charge channel $F^{\rm c}(i\omega_{n},i\omega_{n'},i\nu_{m})$ in the $n-n'$ plane calculated with the QMC as the impurity solver~\cite{RevModPhys.90.025003,PhysRevB.94.235108}. 
As we can see from this figure,
$F^{\rm c}(i\omega_{n},i\omega_{n'},i\nu_{m})$
takes large values in the vicinity of the diagonal line in the $n-n'$ plane. 
${\rm \overline{ph}}$ and pp channels exhibit large values in the vicinity of $\omega_{n}-\omega_{n'}=0$ and $\omega_{n}+\omega_{n'}+\nu_{m}=0$, respectively.
These large values on the diagonal lines come from these channels. 
ph channel takes large values near $\nu_{m}=0$ although it is not depicted in  Fig.~\ref{fig:2020-06-14-00-39}. 
This structure coming from ph, ${\rm \overline{ph}}$, and pp channels is called ``diagonal structure''~\cite{PhysRevB.96.035114,PhysRevB.97.235140,PhysRevB.100.075119,PhysRevB.102.085106}. 
We can also see that
$F^{\rm c}(i\omega_{n},i\omega_{n'},i\nu_{m})$
takes large values in the vicinity of $\omega_{n}=0$ and $\omega_{n'}=0$ lines. 
This +shaped structure is called ``cross structure''~\cite{PhysRevB.96.035114,PhysRevB.97.235140,PhysRevB.100.075119,PhysRevB.102.085106}. 
There is one more characteristic structure which $F^{\rm c}(i\omega_{n},i\omega_{n'},i\nu_{m})$ depends on $\omega_{n}$ and $\omega_{n'}$ independently and has large values near the center of $n-n'$ plane.
We call this ``central structure''.
The contribution from the cross and central structures is important in the strongly correlated regime,
since these two structures come from the higher order diagrams than that of the diagonal structure~[see Appendix~\ref{sec:2021-04-21-23-54} for details].

\begin{figure}[]
  \centering
  {\includegraphics[width=80mm,clip]{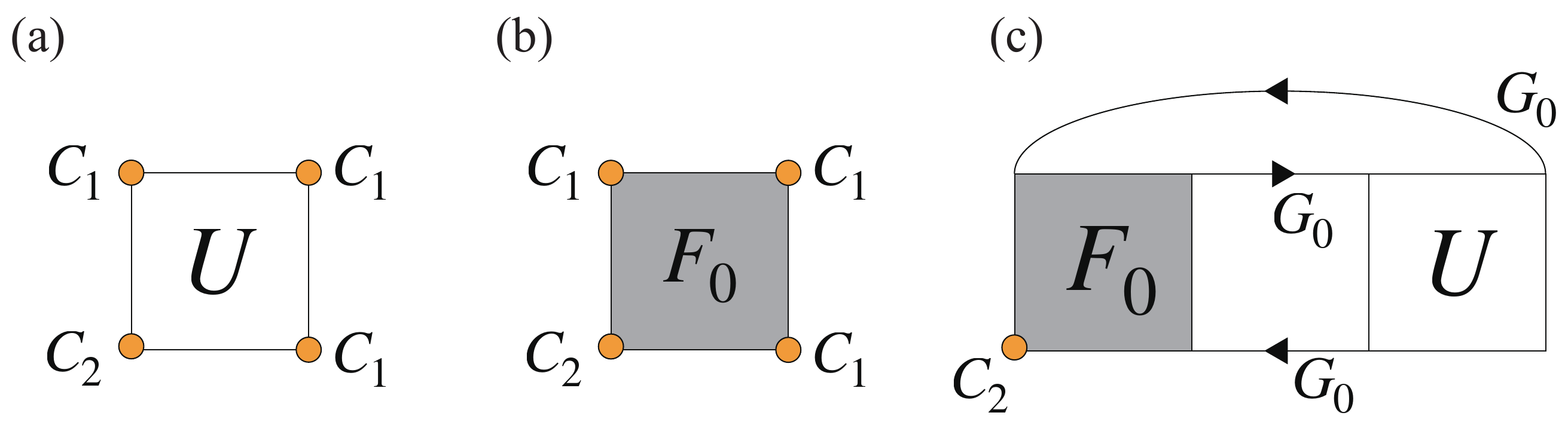}}
  \caption{
    Diagrammatic representation of 
    (a)the full vertex of IPT, 
    (b)the full vertex of IPT+parquet,
    and 
    (c)the self energy, 
  } 
  \label{fig:2020-06-14-11-26}
\end{figure}

\begin{figure}[]
  \centering
  {\includegraphics[width=90mm,clip]{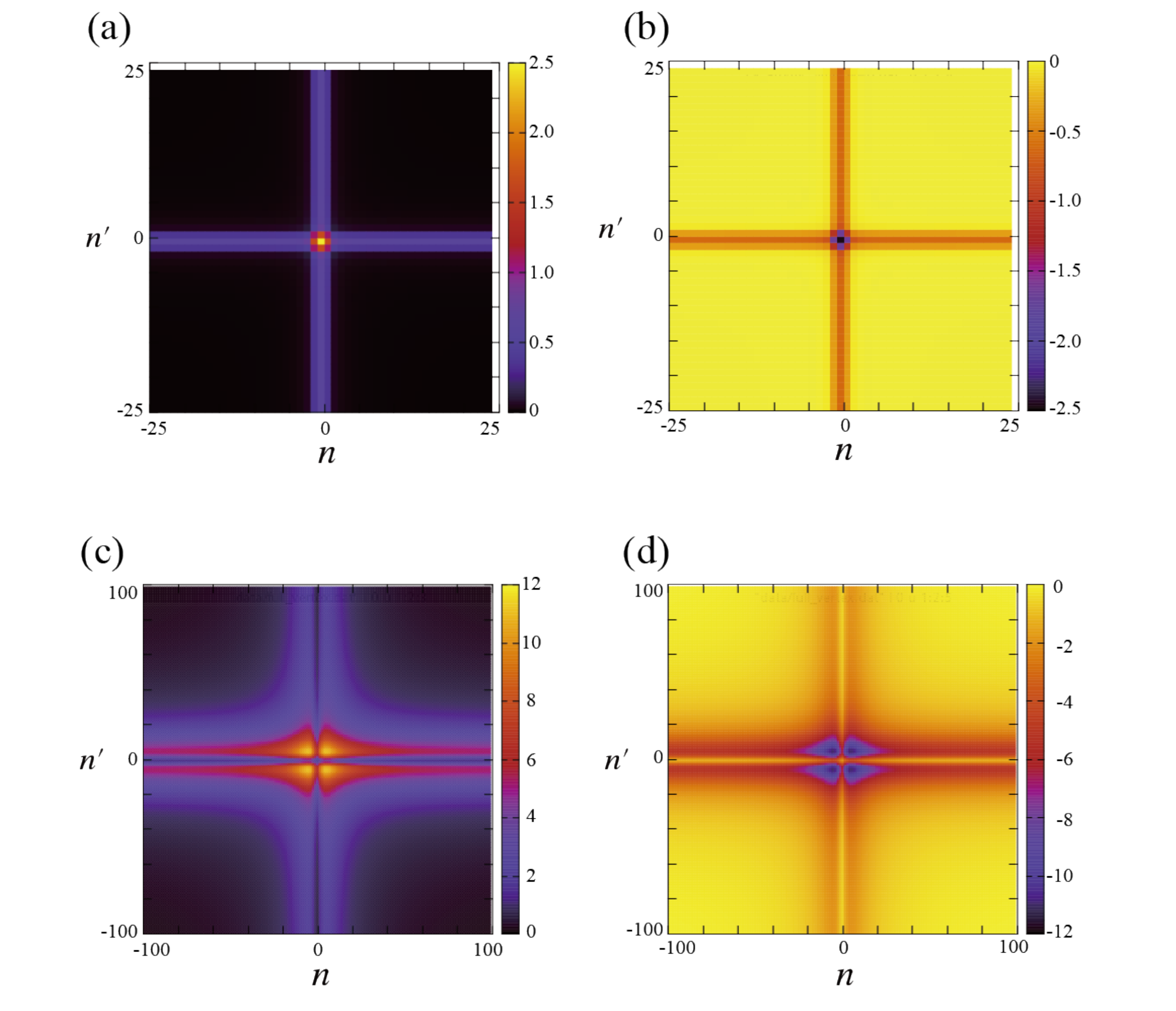}} 
  \caption{The frequency dependence of the full vertex at $\nu_{m}=0$. 
    Upper panels: The frequency dependence of the full vertex in the atomic limit subtracted by the bare interaction and the terms that give the diagonal structure~(${\rm ph,\overline{ph},pp}$ term).
    Lower panels: The frequency dependence of the full vertex in IPT. 
    (a),(c) correspond to the charge channel, and (b),(d) the spin channel.
  } 
  \label{fig:2020-06-03-19-27}
\end{figure}

Given this, 
we move on to IPT. 
Comparing Eq.~(\ref{eq:2020-06-02-20-56}) with the exact expression of the correlation part of the self energy $\hat{\Sigma}^{\rm CR}(i\omega_{n})$ using the full vertex: 
\begin{align}
  {\Sigma}_{\alpha\beta}^{\rm CR}&(\omega_{n}) 
  \nonumber \\
  =&
  T^{2}\sum_{\gamma\lambda}\sum_{\omega_{n'}\nu_{m}} [\hat{F}(\omega_{n},\omega_{n'},\nu_{m})\hat{\chi_{0}}(\omega_{n'},\nu_{m})\hat{U}]_{\alpha\gamma\beta\lambda}
  G_{\gamma\lambda}(\omega_{n}+\nu_{m}) ,
  \label{eq:2020-06-02-20-53}\\
  &\chi_{0,\alpha\beta\gamma\lambda}(\omega_{n},\nu_{m}) = -G_{\alpha\gamma}(\omega_{n})G_{\lambda\beta}(\omega_{n}+\nu_{m})
  \label{eq:2021-06-29-00-56}
\end{align}
the full vertex in IPT can be written as 
\begin{align} 
  [&F_{\rm IPT}(\omega_{n},\omega_{n'},\nu_{m}) ]_{\alpha\beta\gamma\lambda}  
  \nonumber \\
  &=
  C_{2,\alpha\alpha'}(\omega_{n})
  C_{1,\beta\beta'}(\omega_{n}+\nu_{m})
  U_{\alpha'\beta'\gamma'\lambda'}
  C_{1,\gamma'\gamma}(\omega_{n'})
  C_{1,\lambda'\lambda}(\omega_{n'}+\nu_{m})
  \label{eq:2020-06-02-20-42} \\
  &\hspace{20pt} \hat{C}_{1}(\omega_{n}) = \hat{G}_{0}(\omega_{n})\hat{G}^{-1}(\omega_{n}) \label{eq:2020-06-03-19-45} \\
  &\hspace{20pt} \hat{C}_{2}(\omega_{n}) = [\hat{I} - \hat{B}\hat{\Sigma}^{(2)}(\omega_{n})]^{-1}\hat{A} \label{eq:2020-06-03-19-46} .
\end{align} 
The diagrammatic representation of $\hat{F}_{\rm IPT}$ is shown in Fig.~\ref{fig:2020-06-14-11-26}~(a). 
Fig.~\ref{fig:2020-06-03-19-27}~(a)-(b) show the 
the full vertex in the atomic limit $\hat{F}_{\rm atom}$~\cite{PhysRevB.86.125114} subtracted by the terms which give the diagonal structure~(ph,~$\overline{\rm ph}$,~pp terms) 
and constant $\hat{U}$,
and 
(c)-(d) show $\hat{F}_{\rm IPT}$
in a single-band case.
We can see that these structures resemble each other
and hence
we can say that the IPT is an approximation which captures the strong correlation effects by the ``pseudo'' cross and central structures.
On the other hand, 
IPT fails to capture the diagonal structure as we can see from Eq.~(\ref{eq:2020-06-02-20-42}), where the $\omega_{n}-\omega_{n'}$ and $\omega_{n}+\omega_{n'}+\nu_{m}$ dependences are absent. 

This is a new interpretation of IPT and is completely different from the conventional one in which IPT is considered as an interpolation method from the weak and strong correlation limits.
It should be noted that correctly reproducing the frequency dependence of one-body quantities (e.g. self energy) does not necessarily imply reproducing that of two-body quantities. 
Therefore, 
we believe it is more appropriate to recognize that IPT correctly captures the strong correlation effects because it mimics the cross and central structures of the full vertex, 
not simply because it reproduces the exact solution of the self energy in the strong correlation limit.

This re-interpretation naturally leads to the extension of IPT in Sec.~\ref{sec:2020-11-06-17-44}.
As mentioned above,
IPT fails to capture the diagonal structure of the full vertex. 
This structure can be captured by the following replacement. 
\begin{align}
  [&F_{\rm IPT}(\omega_{n},\omega_{n'},\nu_{m}) ]_{\alpha\beta\gamma\lambda}  
  \nonumber \\
  &=
  C_{2,\alpha\alpha'}(\omega_{n})
  C_{1,\beta\beta'}(\omega_{n}+\nu_{m})
  U_{\alpha'\beta'\gamma'\lambda'}
  C_{1,\gamma'\gamma}(\omega_{n'})
  C_{1,\lambda'\lambda}(\omega_{n'}+\nu_{m})
  \label{eq:2020-08-28-01-08} \\
  & \hspace{30pt} \downarrow \nonumber \\
  [&F_{\rm IPT+parquet}(\omega_{n},\omega_{n'},\nu_{m}) ]_{\alpha\beta\gamma\lambda} 
  \nonumber \\
  &=
  C_{2,\alpha\alpha'}(\omega_{n})
  C_{1,\beta\beta'}(\omega_{n}+\nu_{m}) 
  \nonumber \\
  &\times F_{0,\alpha'\beta'\gamma'\lambda'}(\omega_{n},\omega_{n'},\nu_{m})
  C_{1,\gamma'\gamma}(\omega_{n'})
  C_{1,\lambda'\lambda}(\omega_{n'}+\nu_{m}),
  \label{eq:2020-08-28-01-09} 
\end{align} 
where $\hat{C}_{2}$ in Eq.~(\ref{eq:2020-06-03-19-46}) is also replaced with
\begin{align}
  \hat{C}_{2}(\omega_{n}) =& [\hat{I} - \hat{B}\hat{\Sigma}^{\rm CR}_{0}(\omega_{n})]^{-1}\hat{A} .
  \label{eq:2021-06-29-01-20}
\end{align}
The diagrammatic representation of the self energy and full vertices are shown in Fig.~\ref{fig:2020-06-14-11-26}.
Substituting the full vertex in Eq.~(\ref{eq:2020-08-28-01-09}) into the exact expression of the self energy in Eq.~(\ref{eq:2020-06-02-20-53}), 
we obtain Eqs.~(\ref{eq:2020-08-28-02-01})-(\ref{eq:2020-08-28-01-28}).
As mentioned in Sec.~\ref{sec:2020-11-06-17-44},
we employ the simplified parquet method developed in Ref.~\cite{doi:10.1143/JPSJ.79.094707} to obtain $\hat{F}_{0}$.
The approximate full vertex $\hat{F}_{0}$ needs to 
(i)~have only diagonal and constant terms because the cross and central terms are given by $C_{1}$, 
and 
(ii)~be obtained with low numerical cost so as not to lose the advantage of IPT.
The simplified parquet method can meet these requirements because it can provide the diagonal and constant terms with low numerical cost while ignoring the cross and central structures~[see Appendix~\ref{sec:2020-10-09-00-10}].

While we employ simplified parquet method here, 
$\hat{F}_{0}$ can be evaluated by other methods as long as they estimate only the diagonal and constant parts of the full vertex.  
Indeed, in Sec.~\ref{sec:2020-11-08-15-30}, 
we show the result of ``IPT+FLEX''  in which we obtain $\hat{F}_{0}$ in Eq.~(\ref{eq:2020-08-28-01-28}) by the fluctuation exchange~(FLEX) approximation~\cite{PhysRevLett.62.961} for comparison. 
On the other hand, for example, 
the exact full vertex of the atomic limit $\hat{F}_{\rm atom}$ or the full vertex obtained from the non-simplified parquet method are not suitable for $\hat{F}_{0}$ 
since they already have the cross and central structures and hence yield double counting if adopted.

Finally, we should note that the ansatz in Eqs.~(\ref{eq:2020-06-03-19-45}), (\ref{eq:2020-06-03-19-46}) and (\ref{eq:2021-06-29-01-20})~[more generally the choice of $\hat{C}_{1}\neq \hat{C}_{2}$] breaks the crossing symmetry of the full vertex.
In fact, in a separate publication Ref.~\cite{mizuno_s2fedf},
we develop a method in which we define a different full vertex that reproduces the same self energy without breaking the crossing symmetry.


\subsection{How to determine the parameters $A$, $B$, and $\mu_{0}$}\label{sec:2021-06-24-14-06}
The remaining problem here is how to deal with the parameters $\hat{A}$, $\hat{B}$ in Eq.~(\ref{eq:2020-08-28-02-01}), and $\hat{\mu}_{0}$ in Eq.~(\ref{eq:2020-07-02-19-01}) in the multi-band systems.
In MO-IPT~\cite{PhysRevLett.91.156402,Dasari2016},
the two or more particle effects are added in the form of the static correlation functions 
when the single-orbital IPT is extended to the multi-orbital one.
By contrast,
in our formalism,
these effects are already considered in the form of the diagonal terms of the dynamical full vertex  obtained by the parquet equations.
Therefore, we simply extend the single-orbital representation of parameters  Eq.~(\ref{eq:2021-04-02-22-59}) to  multi-orbital forms, 
as follows.
\begin{align}
  A_{\alpha\beta} =& \delta_{\alpha\beta} \label{eq:2020-06-04-16-20} \\
  B_{\alpha\beta} =& \delta_{\alpha\beta} \dfrac{ N_{\rm orbital}^{-1} \sum_{\gamma}U_{\alpha\alpha\gamma\gamma}(1-2n_{\gamma}) + \mu_{0\alpha\alpha}-\mu}{\sum_{\gamma}U_{\alpha\alpha\gamma\gamma}n_{0\gamma}(1-n_{0\gamma})U_{\gamma\gamma\alpha\alpha}},
  \label{eq:2020-06-04-16-19} 
\end{align}
where $n_{\alpha}, n_{0\alpha}$ is the band filling evaluated from $\hat{G}, \hat{G}_{0}$. 
In addition,
we add a degree of freedom to the pseudo chemical potential as $\mu_{0} \to \hat{\mu}_{0}$~[i.e. not a scalar but a diagonal matrix] 
in order to satisfy the condition $n_{\alpha}=n_{0\alpha}$, 
and so $\hat{A}$ is fixed to unity.
This condition is needed for the following reason. 
According to the interpretation introduced in Sec.~\ref{sec:2021-04-22-01-15},
the correction factor $\hat{C}_{1}=\hat{G}_{0}\hat{G}^{-1}$ captures the strong correlation effects.
Hence, $\hat{C}_{1}$ needs to increase in the appropriate regions of filling, 
namely, 
in the vicinity of half filling. 
In the multi-band case,
if the pseudo chemical potential is a single scalar parameter $\mu_{0}$~[i.e. independent of the band index $\alpha$], 
$n_{0\alpha}$ can be different from $n_{\alpha}$ in general even if $\sum_{\alpha}n_{0\alpha}=\sum_{\alpha}n_{\alpha}$ is satisfied.
For example, it is possible that $n_{0\alpha}$ is at half filling but $n_{\alpha}$ is away from it, or vice versa.
$C_{1}$ is not appropriately given under these circumstances.
Therefore, 
we need to regard the pseudo chemical potential as a diagonal matrix $\hat{\mu}_{0}$ by adding the band index $\alpha$ to $\mu_{0}$ for the condition $n_{\alpha}=n_{0\alpha}$.

Also, we should note that the first term in the numerator in Eq.(\ref{eq:2020-06-04-16-19}) is divided by only the number of orbitals $N_{\rm orbital}$ [i.e., not by the number of sites $N_{\rm site}$]. 
This is because we assume the interaction which does not have site-off-diagonal elements but 
has the orbital-off-diagonal elements that is comparable with the diagonal elements in magnitude.
This interaction is valid in many realistic systems.
The parameter $\hat{B}$ is related to the electron-hole asymmetry [see Appendix.~\ref{sec:2021-01-06-00-42} in detail]. 
If we take the summation over orbital index without $N_{\rm orbital}^{-1}$, 
$\hat{B}$ is overestimated and then the electron-hole asymmetry is overestimated.
On the other hand,
if we divide the first term in the numerator in Eq.(\ref{eq:2020-06-04-16-19}) by $N_{\rm site}$, 
$\hat{B}$ is underestimated since the summation over the site-off-diagonal elements is zero.
When we consider the interaction which has the site-off-diagonal elements or does not have the orbital-off-diagonal elements, 
the expression of $\hat{B}$ in Eq.(\ref{eq:2020-06-04-16-19}) have to be modified.

\section{Results}\label{sec:2020-12-25-04-05}
In this section,
we show the results of IPT+parquet method.
We use the quasi-particle weight as a probe of the correlation effects.
The quasi-particle weight is defined as 
\begin{align}
  Z_{\alpha} 
  =&
  \left(1-\dfrac{{\rm Im}\Sigma_{\alpha\alpha}(\omega_{n})}{\omega_{n}}\Bigl|_{\omega_{n}\to 0} \right)^{-1}.
  \label{eq:2020-07-02-16-26}
\end{align}
$Z_{\alpha}$ is roughly proportional to the inverse of the effective mass, and $Z_{\alpha}=0$ corresponds to the insulating state.
In this study, however, 
we adopt the following definition instead of Eq.~(\ref{eq:2020-07-02-16-26}) for calculational simplicity.
\begin{align} 
  Z_{\alpha} 
  =&
  \left(1-\dfrac{{\rm Im}\Sigma_{\alpha\alpha}(\omega_{n})}{\omega_{n}}\Bigl|_{n=0}\right)^{-1}.
  \label{eq:2020-07-02-16-27}
\end{align}
Also,
in this study, we adopt the definition of the band-filling $n_{\alpha} \ (=T\sum_{n}G_{\alpha\alpha}(i\omega_{n})e^{-i\omega_{n}0})$ as the number of electrons per site per spin.

\subsection{Single-orbital model}\label{sec:2020-11-08-15-30}
\begin{figure}[]
  \centering
  {\includegraphics[width=85mm,clip]{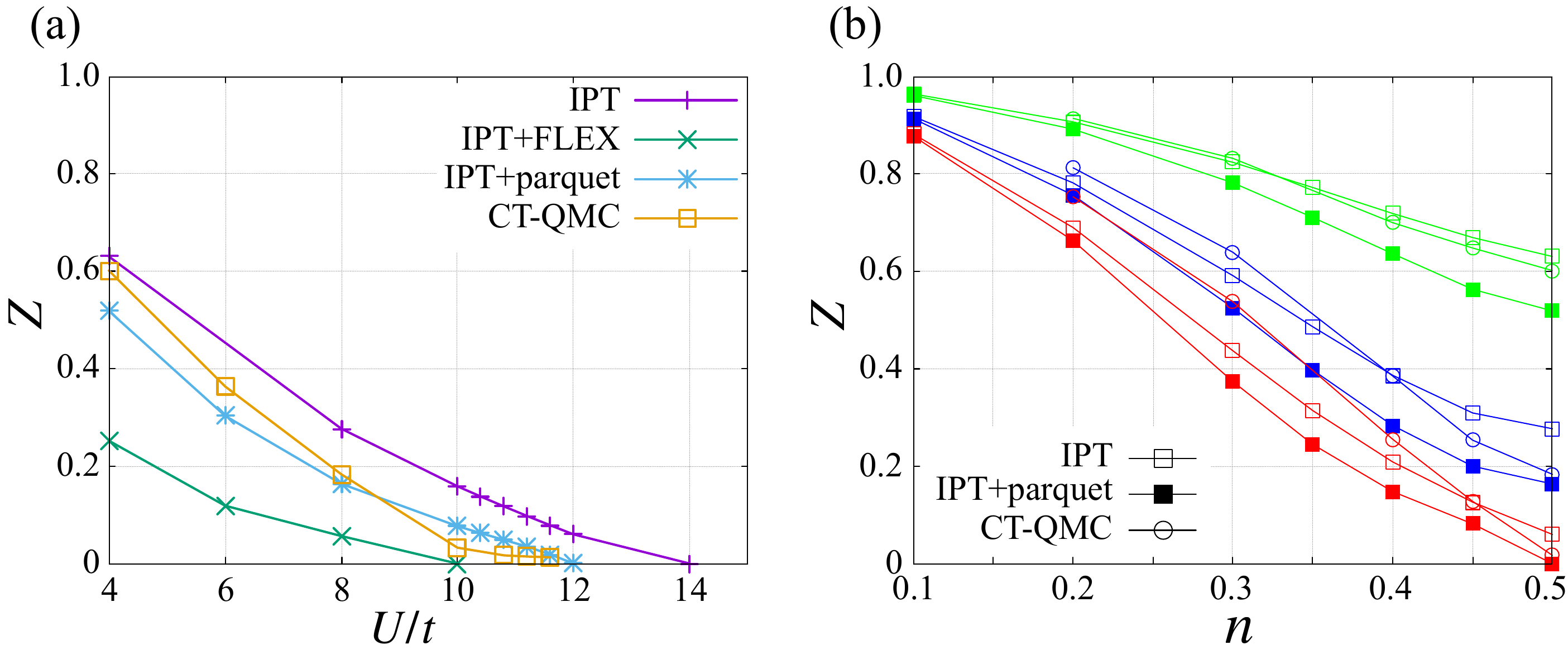}} 
  \caption{Comparison of quasi-particle weight among methods.
    (a)~The violet, blue, green and yellow lines indicate $Z$ at half filling obtained by IPT, IPT+FLEX, IPT+parquet and CT-QMC, respectively.
    (b)~The green, blue and red lines indicate $Z$ for $U/t=4,8,12$, respectively. The open square corresponds to modified-IPT, solid square IPT+parquet, and the circle CT-QMC.   
    The temperature is $T/t=0.04$ in both figures.
  } 
  \label{fig:2020-06-30-17-36}
\end{figure}

We study the square lattice model as a benchmark in the single-orbital systems.
We set the temperature $T/t=0.04$
and 
we take $64\times 64$ $k$-meshes and 4096 Matsubara frequencies,
where $t$ is the nearest neighbor hopping.
Figure~\ref{fig:2020-06-30-17-36} shows the quasi-particle weight calculated by several methods as a function of (a) the interaction $U$ and (b) the band-filling $n$.
IPT+FLEX is the method in which  $F_{0}$ in Sec.~\ref{sec:2020-11-06-17-44} is obtained by the fluctuation exchange~(FLEX) approximation~\cite{PhysRevLett.62.961}.
In CT-QMC calculation, we use the CTHYB~\cite{Seth2016274,PhysRevLett.97.076405,PhysRevB.74.155107,Gull_phdthesis,lewin_thesis,triqs_ctqmc_solver_legendre} code based on  the TRIQS library\cite{triqs}. 
We find that 
the result of IPT+parquet is the closest to that of the numerically exact CT-QMC. 
On the other hand,
$Z$ is overestimated in IPT and underestimated in IPT+FLEX, 
since the two-particle fluctuations are underestimated in IPT and overestimated in IPT+FLEX. 
This quantitative improvement from (conventional) IPT is purely due to adding the diagonal terms estimated by parquet equations
since the modified parameters are the same ($A=1, B=0$) in both IPT and IPT+parquet.
In Fig.~\ref{fig:2020-06-30-17-36}~(b),
$Z$'s calculated by IPT, IPT+parquet, and CT-QMC are plotted as functions of the band filling $n$ for $U/t=4,8,12$.
In the region away from half-filling, IPT+parquet tends to underestimate $Z$ (overestimate the correlation effect) compared to IPT.


\subsection{Two-orbital model}
\begin{figure}[h]
  \centering
  {\includegraphics[width=85mm,clip]{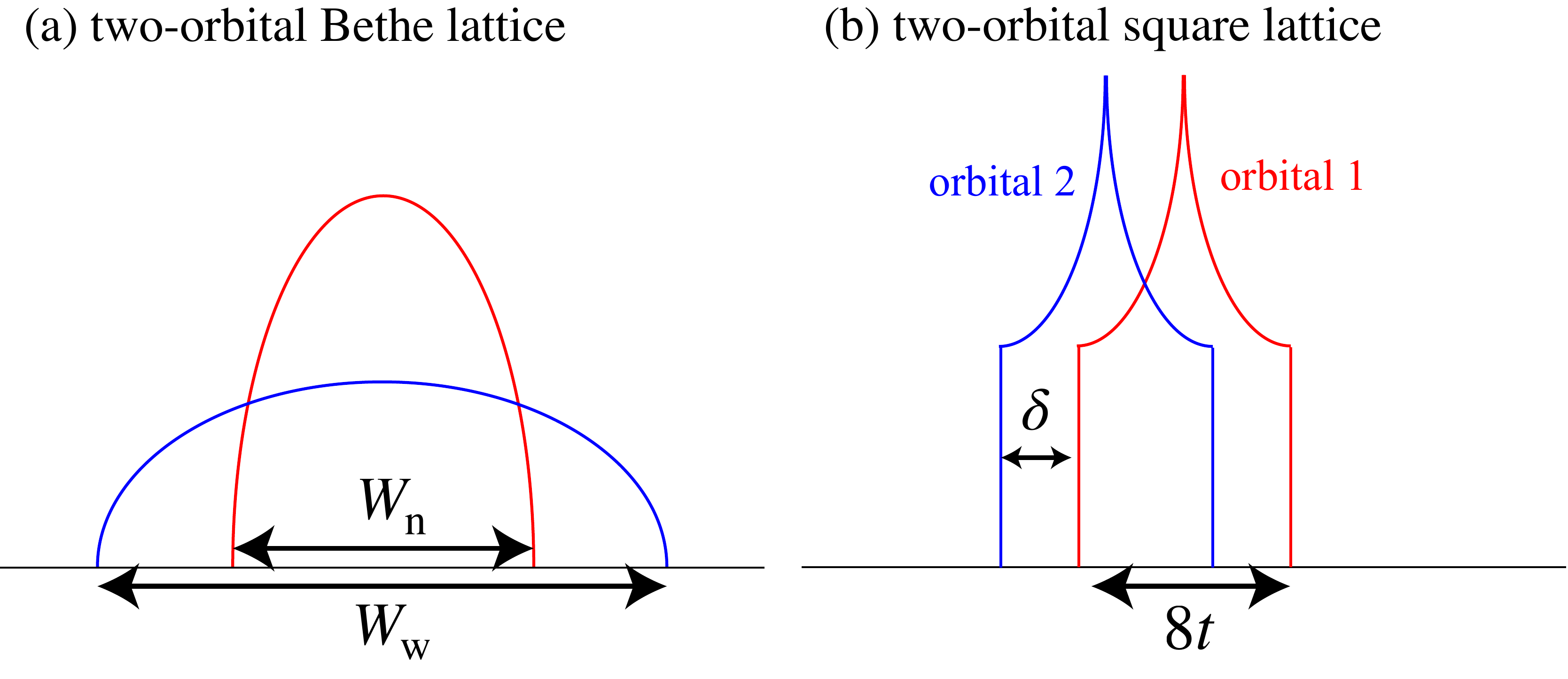}} 
  \caption{
    The non-interacting density of states of the models: 
    Left and right panels show the density of states of 
    two-orbital Bethe lattice,
    two-orbital square lattice,
    respectively.
  } 
  \label{fig:2020-08-13-15-24}
\end{figure}

Here, we study the two-orbital (single-site) model.
The one body part of the Hamiltonian is expressed as 
\begin{align}
  H_{0} 
  =&
  \sum_{ij}\sum_{\alpha\beta}t_{ij,\alpha\beta}c^{\dagger}_{i\alpha}c_{j\beta} - \mu \sum_{i}\sum_{\alpha}n_{i\alpha}.
  \label{eq:2020-07-12-20-15}
\end{align}
The interaction part of the Hamiltonian is expressed as 
\begin{align}
  H_{\rm int}
  =&
  \sum_{l} U n_{l \uparrow}n_{l\downarrow} 
  +
  \sum_{l_{1}\neq l_{2}} \sum_{\sigma_{1}\sigma_{2}} U' n_{l_{1}\sigma_{1}} n_{l_{2}\sigma_{2}}
  \nonumber \\
  &+
  \sum_{l_{1}l_{2}} J \bm{S}_{l_{1}} \cdot \bm{S}_{l_{2}} 
  +
  \sum_{l_{1}l_{2}} J' c^{\dagger}_{l_{1}\uparrow}c^{\dagger}_{l_{2}\downarrow} c_{l_{2}\downarrow}c_{l_{2}\uparrow},
  \label{eq:2020-07-02-21-25}
\end{align}   
where 
the degrees of freedom of orbital are expressed by $l$ and spin by $\sigma$.
$U^{(\prime)}$ is the intra~(inter)-orbital interaction, 
and 
$J$ and $J'$ represent the Hund's coupling and pair hopping, respectively. 
Then, 
the interaction matrices in the charge and spin channels are expressed as
\begin{align}
  \Bigl( U^{\rm c}_{l_{1}l_{2}l_{3}l_{4}}, U^{\rm s}_{l_{1}l_{2}l_{3}l_{4}} \Bigr)
  =
  \begin{cases}
    (U, U) \hspace{20pt} &(l_{1}=l_{2}=l_{3}=l_{4}) \\
    (2U'-J, J) \hspace{20pt} &(l_{1}=l_{2}\neq l_{3}=l_{4}) \\
    (2J-U', U') \hspace{20pt} &(l_{1}=l_{3}\neq l_{2}=l_{4}) \\
    (J', J') \hspace{20pt} &(l_{1}=l_{4}\neq l_{2}=l_{3})
  \end{cases}.
  \label{eq:2020-07-02-21-39}
\end{align}
Figure~\ref{fig:2020-08-13-15-24} shows the non-interacting density of states of the models which we study here.

\subsubsection{Two-orbital Bethe lattice}\label{sec:2021-01-06-01-35}

\begin{figure}[h]
  \vspace{10pt}
  \centering
  {\includegraphics[width=85mm,clip]{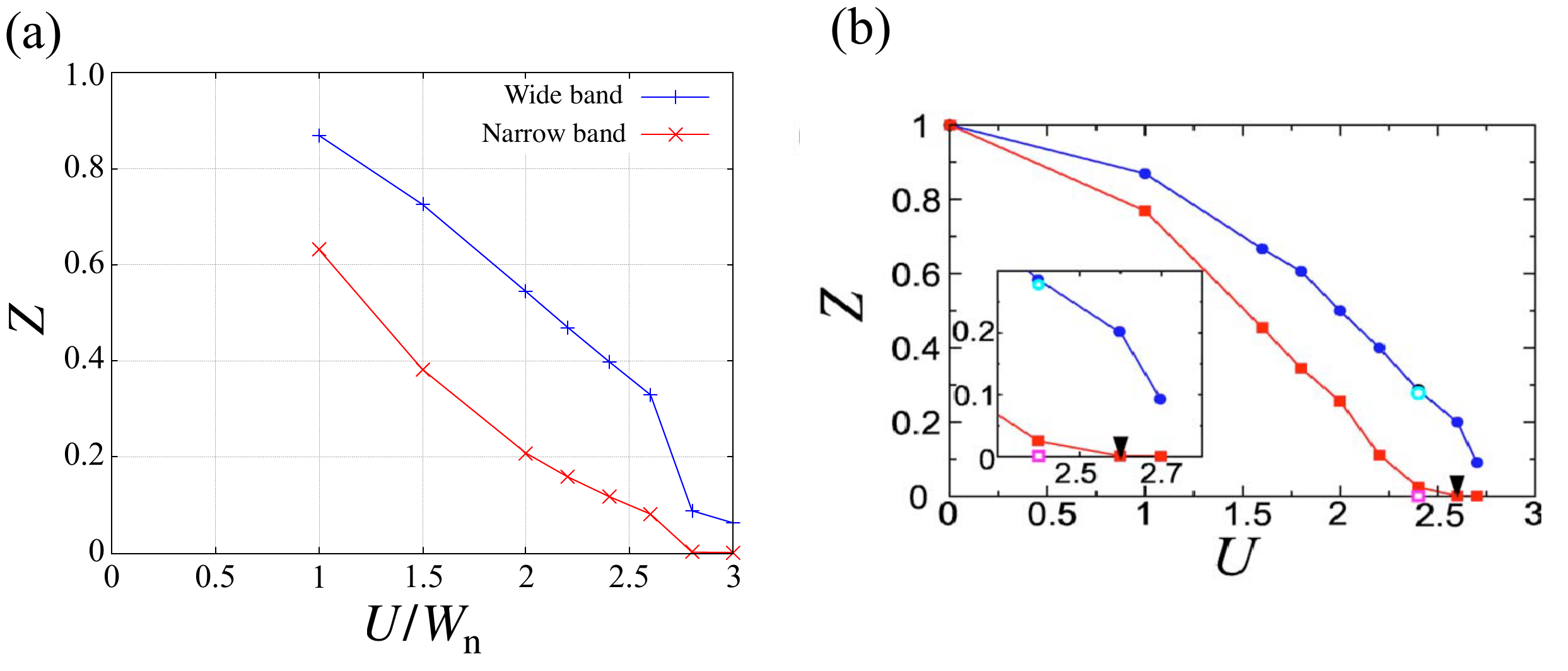}} 
  \caption{Quasi-particle weight $Z$ of the two-orbital Bethe lattice as a function of the interaction $U$.
    The temperature is $T/W_{\rm n}=0.02$.
    The result of IPT+parquet is shown in (a), and projective-QMC in (b)[This figure is taken from Ref.~\cite{PhysRevB.72.201102}].
    Red and blue lines indicate the narrow band and wide band, respectively.
  } 
  \label{fig:2020-04-29-15-32}
\end{figure}

We consider the two-orbital Bethe lattice model in which two bands with different band width exist.
We set $W_{\rm w}/W_{\rm n}=2$, where $W_{\rm n}$ and $W_{\rm w}$ represent the half band width of the narrow and wide bands, respectively.
We also set $U'=U-2J, J=U/4, J'=0$, and the temperature $T/W_{\rm n}=0.02$.
We take 2000 real-frequency meshes and 4096 Matsubara frequencies.
The quasi-particle weight against the interaction $U$ for each orbital is plotted in Fig.~\ref{fig:2020-04-29-15-32}.
(a) is the result of IPT+parquet and (b) is that of the projective-QMC~(PQMC) in Ref.~\cite{PhysRevB.72.201102}. 
We can see a good agreement between the two methods.
The orbital selective Mott transition, in which the energy gap opens in the narrow band whereas the wide band is still metallic, occurs at $U/W_{\rm n}\sim 2.7$.  

\subsubsection{Two-orbital square lattice}\label{sec:2020-12-20-13-04}

\begin{figure}[]
  \centering
  {\includegraphics[width=55mm,clip]{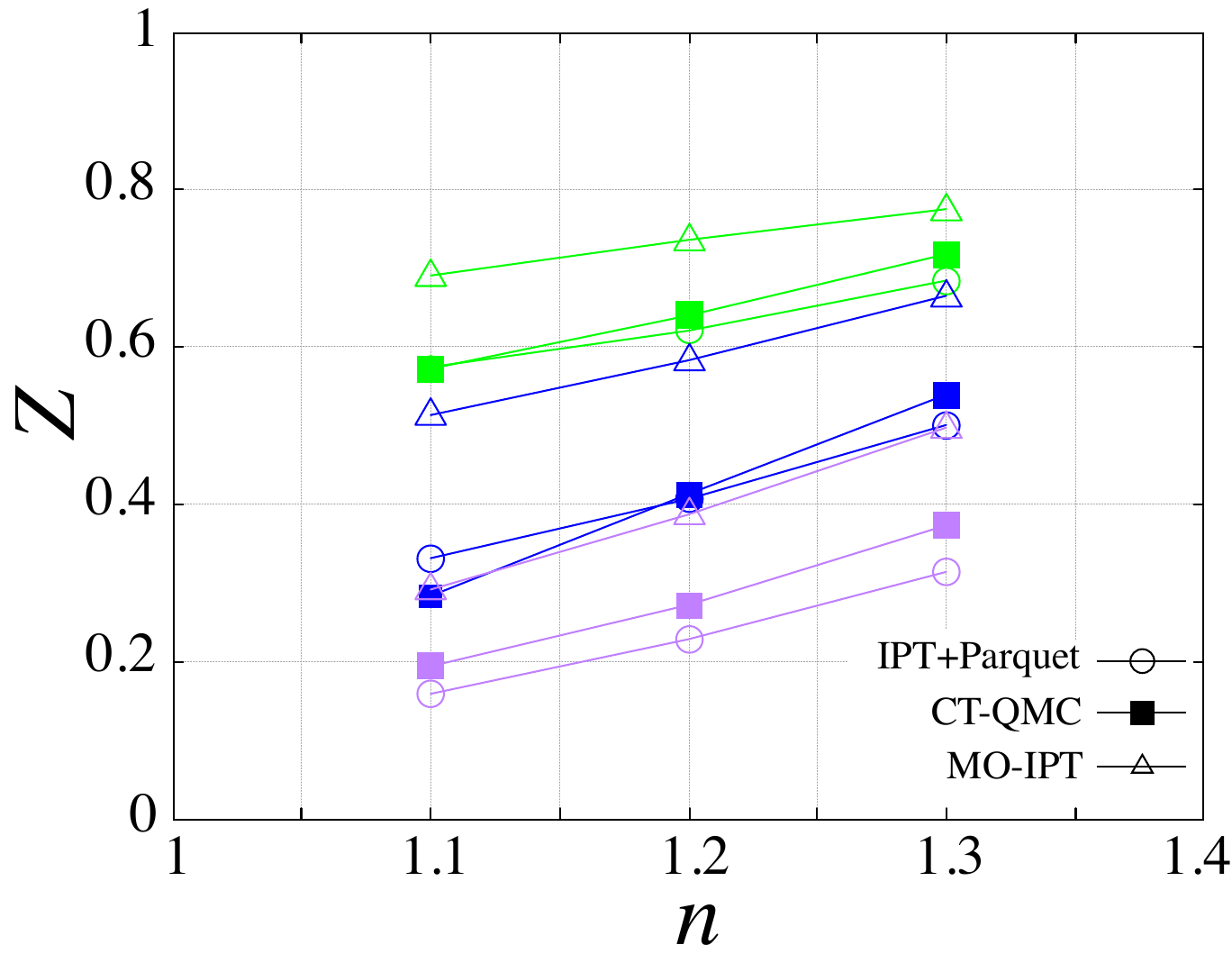}} 
  \caption{Quasi-particle weight $Z$ of the two-orbital square lattice as a function of the band filling $n$.
    The temperature is $T/t=0.2$ and the onsite energy difference is $\delta/t = 0$.
    Green, blue, and purple lines indicate the results at $U/t=4,6$, and $10$, respectively.
    Circle, square, and triangle represent IPT+parquet, CT-QMC, and MO-IPT, respectively.
  } 
  \label{fig:2021-01-07-14-51}
\end{figure}%

\begin{figure}[]
  \centering
  {\includegraphics[width=80mm,clip]{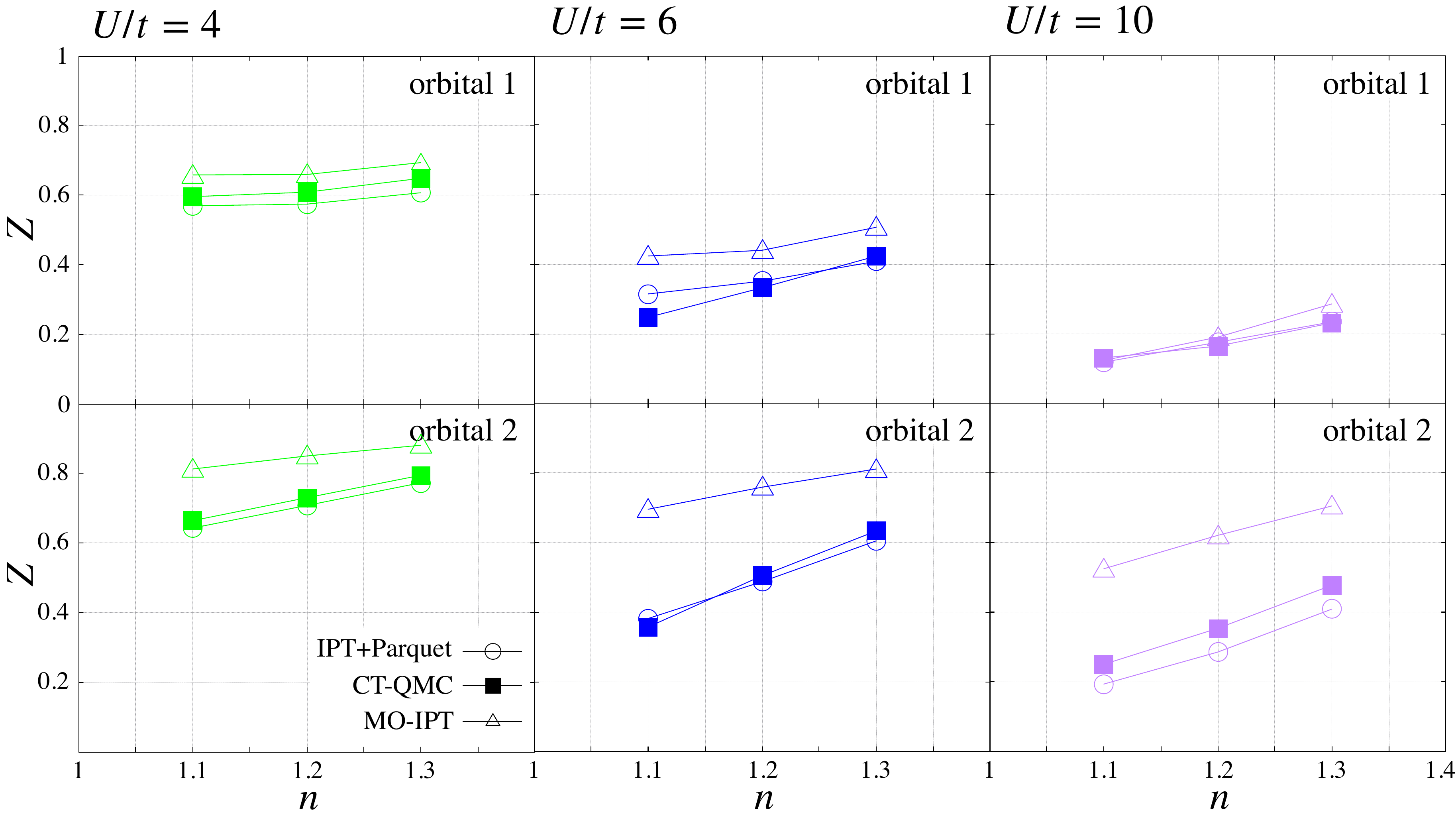}} 
  \caption{Quasi-particle weight $Z$ of the two-orbital square lattice as a function of the band filling $n$.
    The temperature is $T/t=0.2$ and the onsite energy difference is $\delta/t = 1.6$.
    Green, blue, and purple lines indicate the results at $U/t=4,6$, and $10$, respectively.
    Circle, square, and triangle represent IPT+parquet, CT-QMC, and MO-IPT, respectively.
  }
  \label{fig:2020-08-02-21-19}
\end{figure}%

We study the two-orbital square lattice model which has only the intra-orbital nearest neighbor hopping. 
Here, 
we compare three impurity solvers: IPT+parquet, CT-QMC, and MO-IPT. 
In the CT-QMC calculation, we use the CTHYB package~\cite{Seth2016274,PhysRevLett.97.076405,PhysRevB.74.155107,Gull_phdthesis,lewin_thesis,triqs_ctqmc_solver_legendre} based on  the TRIQS library~\cite{triqs}. 
In MO-IPT calculation, the spin-flip and the pair-hopping processes are ignored since MO-IPT supports only the density-density type interactions~\cite{Dasari2016}. 
We set $t_{1}=t_{2}=t$, where $t_{\alpha}=t_{i,i+1, \alpha\alpha}$ is the nearest neighbor hopping of orbital $\alpha$ and $t$ is the unit of energy.
The onsite energy difference $\delta=t_{ii,11}-t_{ii,22}$, 
and 
the interactions $U'=U-2J, J=J'=U/4$.
We take $32\times 32$ $k$-meshes and 4096 Matsubara frequencies 
and we fix the temperature $T/t=0.2$.
Here, we have intentionally omitted the calculation result for half-filling, 
which turns out to require special care due to spontaneous symmetry breaking.
This point will be studied in detail in a separate publication.

We start with  the $\delta=0$ case (orbital degenerate case).
Figure~\ref{fig:2021-01-07-14-51} shows the quasi-particle weight $Z$ obtained by  three methods MO-IPT, IPT+parquet, and CT-QMC as a function of the filling $n$ for several interaction strength $U$ 
at $\delta=0$.
Since the two orbitals are equivalent at $\delta=0$,
we show only $Z$ of orbital 1 and omit the orbital index.
The MO-IPT results significantly deviate from those of CT-QMC, which is qualitatively consistent with the situation for the two-orbital Bethe lattice model in Ref.~\cite{Dasari2016}.
By contrast,
$Z$ of IPT+parquet agrees well with that of CT-QMC. 

\begin{figure}[]
  \centering
  {\includegraphics[width=75mm,clip]{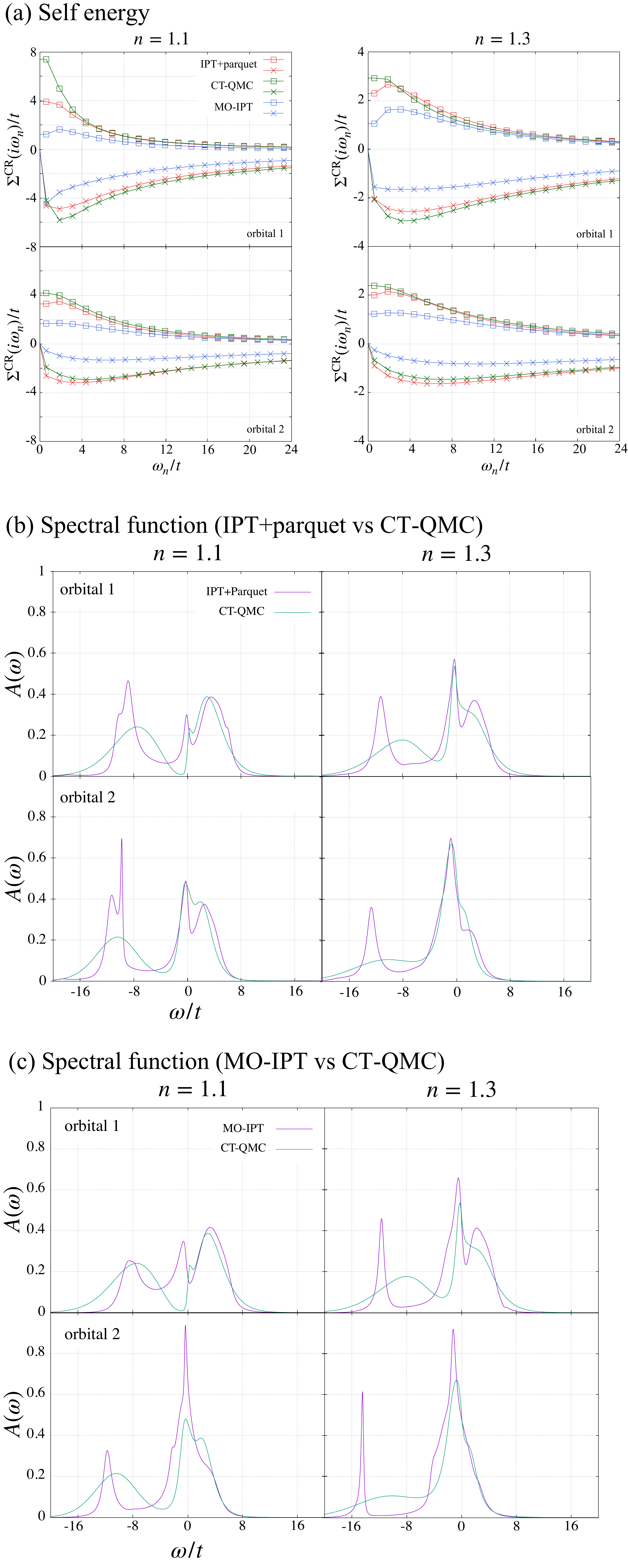}} 
  \caption{
    (a)~The correlation part of the self energy $\Sigma^{\rm CR}(\omega)$ of the two-orbital square lattice.
    $\Sigma^{\rm CR}(i\omega_{n})$  of orbital 1 is shown in upper panel and orbital 2 in lower panel.
    Square and cross symbols indicate the real and imaginary parts, respectively.
    Red, green, and blue lines indicate $\Sigma^{\rm CR}(i\omega_{n})$ obtained by IPT+parquet, CT-QMC and  MO-IPT, respectively.
    (b)(c)~Spectral function $A(\omega)$ of the two-orbital square lattice for several fillings.
    Interaction strength is $U/t=10$.
    The spectral function $A(\omega)$  of orbital 1 is shown in upper panel and orbital 2 in lower panel.
    Purple lines indicate  $A(\omega)$ obtained by (b)~IPT+parquet and (c)~MO-IPT.
    Green lines indicate $A(\omega)$ obtained by CT-QMC.
    The temperature is $T/t=0.2$
    , and 
    the onsite energy difference is $\delta/t=1.6$.
  } 
  \label{fig:2020-12-21-21-44}
\end{figure}

We move on to the $\delta/t=1.6$ case (orbital non-degenerate case).
Figure~\ref{fig:2020-08-02-21-19} shows the quasi-particle weight $Z$ at $\delta/t=1.6$.
The deviations of $Z$'s of MO-IPT from that of CT-QMC are largely different between the two orbitals
\footnote{
  This is qualitatively consistent with the results of the two-orbital Bethe lattice with crystal field splitting shown in Sec.~3.5 in Ref.~\cite{Dasari2016}.
  The situation studied in Sec.~3.5 in Ref.~\cite{Dasari2016} is even further away from half-filling than  $n=1.3$ studied here. 
  [$n_{\rm tot}=1.1$ in Ref.~\cite{Dasari2016}  corresponds to $n=0.55$ in our study, 
    and
    by electron-hole transformation,  
  this corresponds to $n=1.45$ in our case.]
  The deviations of the MO-IPT results from that of CT-QMC are larger than that in Ref.~\cite{Dasari2016} 
  since the DOS of the two dimensional square lattice exhibits a van Hove singularity (in contrast to the Bethe lattice adopted in Ref.~\cite{Dasari2016}),
  so that the quasi-particle weight is sensitive to the on-site energy difference.
  These results suggest that caution has to be taken when we apply the MO-IPT to systems with on-site energy difference.
} .
IPT+parquet is found to improve the situation.
This improvement comes from adding the degree-of-freedom to the pseudo chemical potential $\mu_{0}$, 
which enables the IPT+parquet method to capture the strong correlation effects more appropriately in both orbitals as explained in Sec.~\ref{sec:2020-11-06-17-44}. 
Figure~\ref{fig:2020-12-21-21-44}~(a) shows the correlation part of the self energy $\Sigma^{\rm CR}(i\omega_{n})$ obtained by these three methods. 
We can see that $\Sigma^{\rm CR}(i\omega_{n})$ of IPT+parquet shows better agreement with CT-QMC than MO-IPT, not only in orbital 2 but also in orbital 1,
and 
not only in the imaginary part which contributes the quasi-particle weight $Z$ but also in the real part.
Figure \ref{fig:2020-12-21-21-44}~(b)-(c) show the spectral function $A(\omega)$ obtained by performing the analytic continuation with Pad${\rm \acute{e}}$ approximation in IPT+parquet and MO-IPT,
and with Maximum entropy method using the $\Omega$Maxent code~\cite{PhysRevE.94.023303} in CT-QMC. 
Similarly to  $\Sigma^{\rm CR}(i\omega_{n})$, 
$A(\omega)$ of IPT+parquet shows better agreement with that of CT-QMC.
Especially at $n=1.1$, 
IPT+parquet shows an improvement from MO-IPT.

%
%

\subsection{Bilayer model}\label{sec:2020-12-21-16-02}

\begin{figure}[]
  \centering
  {\includegraphics[width=50mm,clip]{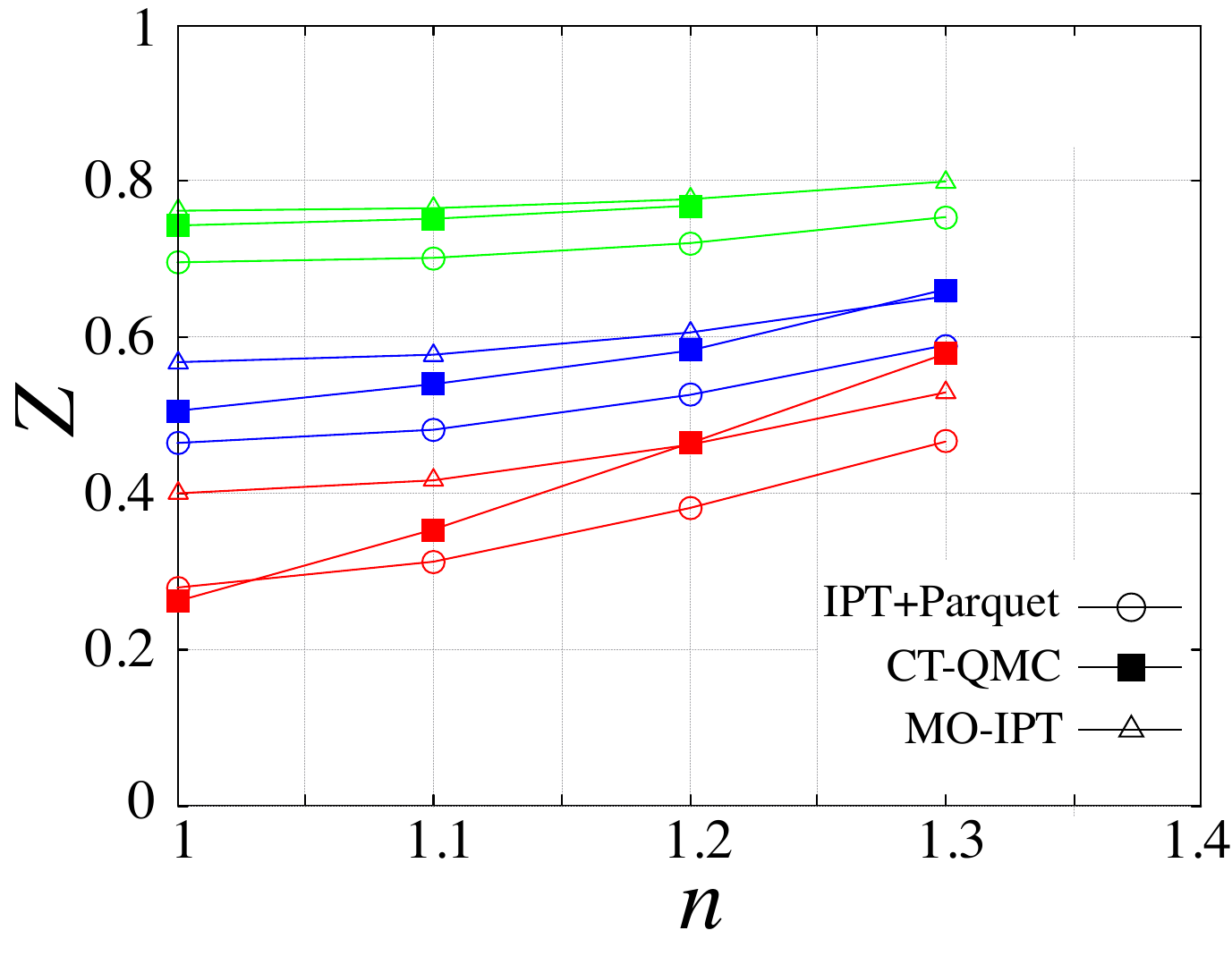}} 
  \caption{Quasi-particle weight $Z$ of the bilayer model as a function of the interaction $n$.
    Green, blue, and red lines indicate the results at $U/t=4,6,8$, respectively.
    Circle, square, and triangle represent IPT+parquet, CT-QMC, and MO-IPT, respectively.
  } 
  \label{fig:2020-11-07-18-45}
\end{figure}

\begin{figure}[]
  \centering
  {\includegraphics[width=85mm,clip]{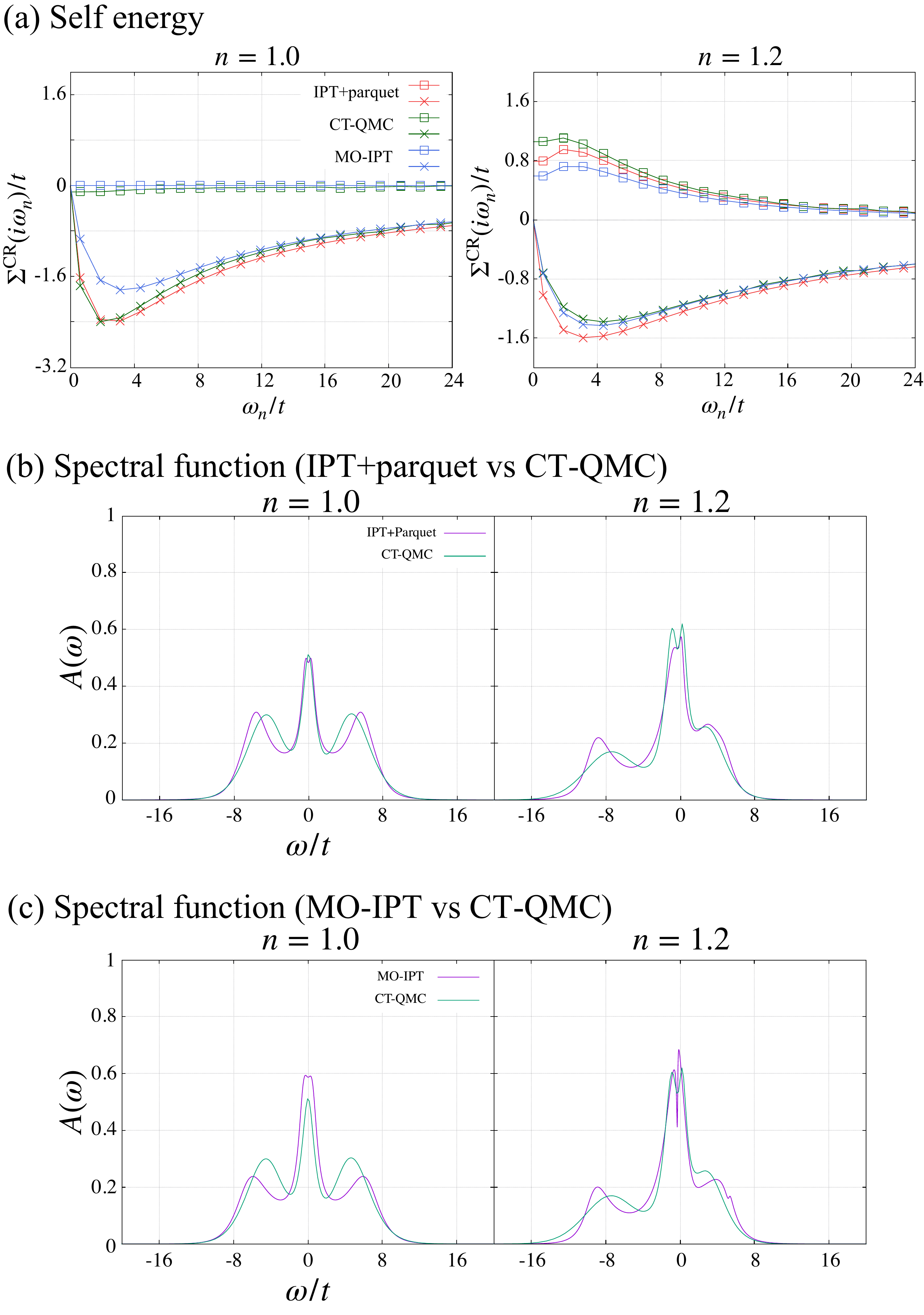}} 
  \caption{
    (a)~The correlation part of the self energy $\Sigma^{\rm CR}(\omega)$ of the bilayer model.
    Square and cross symbols indicate the real and imaginary parts, respectively.
    Red, green, and blue lines indicate  $\Sigma^{\rm CR}(i\omega_{n})$ obtained by IPT+parquet, CT-QMC and  MO-IPT, respectively.
    (b)(c)~Spectral function $A(\omega)$ of the bilayer model for several fillings.
    Purple lines indicate $A(\omega)$ obtained by (b)IPT+parquet and (c)MO-IPT.
    Green lines indicate $A(\omega)$ obtained by CT-QMC.
    Interaction strength is $U/t=8$,
    the temperature $T/t=0.2$
    , and 
    the ratio of hoppings $t_{\perp}/t=1.0$.
  } 
  \label{fig:2020-12-21-21-58}
\end{figure}

Here,
we study the bilayer model on the square lattice 
as a benchmark of systems with multiple sites (with one orbital per site) in a unit cell.
The Hamiltonian of this model is expressed as 
\begin{align}
  H
  =&
  \sum_{\braket{ij}}\sum_{\alpha}t c^{\dagger}_{i\alpha}c_{j\alpha} + \sum_{i}\sum_{\alpha\neq\beta}t_{\perp}c^{\dagger}_{i\alpha}c_{i\beta} 
  +
  \sum_{i}\sum_{\alpha}Un_{i\alpha}n_{i\alpha},
  \label{eq:2020-07-12-21-39}
\end{align}
where $t~(t_{\perp})$ represents the intra~(inter)-layer hopping and $U$ the on-site interaction,
The temperature is fixed as $T/t=0.2$, 
and the hopping ratio $t_{\perp}/t=1.0$.
We take $32\times 32$ $k$-meshes and 4096 Matsubara frequencies.
Since the two sites are equivalent in this model,
we show only the quantities of site 1 and omit the site index.
The quasi-particle weight $Z$ is plotted in Fig.~\ref{fig:2020-11-07-18-45}. 
When $U/t=4$, 
MO-IPT shows better agreement with CT-QMC than IPT+parquet.
When $U$ is increased, 
we can see the tendency that CT-QMC agrees with IPT+parquet~(MO-IPT) near~(away from) half-filling,
similarly to the single-orbital case in Sec.~\ref{sec:2020-11-08-15-30}.
Figure~\ref{fig:2020-12-21-21-58}~(a)
shows the correlation part of the self energy $\Sigma^{\rm CR}(i\omega_{n})$ with $U/t=8$.
As we can also see from $Z$,
IPT+parquet shows better agreement with CT-QMC at $n=1.0$.
At $n=1.2$,
the agreement between MO-IPT and CT-QMC is better in imaginary part 
whereas the agreement between IPT+parquet and CT-QMC is better in real part.
In this model, 
the modified parameters $\mu_{0}, A, B$ are the same between IPT+parquet and MO-IPT 
since the two sites are equivalent and only the onsite interaction is considered~[the interaction matrix has no off-diagonal part in terms of degree-of-freedom].
So the differences come from  the two-particle fluctuation and the off-diagonal part of self energy, 
which are not considered in MO-IPT.
Figure~\ref{fig:2020-12-21-21-58}~(b)-(c)
show the spectral function $A(\omega)$ with $U/t=8$. 
IPT+parquet shows better agreement with CT-QMC at $n=1.0$.
At $n=1.2$, 
CT-QMC and MO-IPT show better agreement in terms of the width of the central peak
whereas 
CT-QMC and IPT+parquet show better agreement in terms of the shape of $A(\omega)$.
This reflects the fact that the imaginary part of the Matsubara self energy is mainly related to the strength of renormalization~[width of the central peak of $A(\omega)$] 
and 
the real part is related to the electron-hole asymmetry.

%

\section{Discussion}\label{sec:2021-04-07-14-03}

\subsection{Validity and Advantage}

Here, we discuss the validity and the advantage of  the IPT+parquet method.
The full vertex in the IPT+parquet method is represented by a simple product of functions as in Eqs.~(\ref{eq:2020-08-28-01-28}) and (\ref{eq:2020-08-28-01-09}). 
This simple product form of the full vertex 
can be justified to some extent  by considering the nature of the two-particle functions,
which are ingredients of the diagrams that give contributions of the cross and central structures.
We can also find that 
the validity of this simple product form of the full vertex 
becomes higher when the contributions of diagrams which give the cross and central structures become larger~[see Appendix~\ref{sec:2021-04-21-23-54} for details].
In the single-orbital case,
indeed,
we can see this simple product form of the full vertex in the atomic limit~[see Appendix~\ref{sec:2021-04-03-00-07}].

Comparing with the existing IPT formalisms, the IPT+parquet method has the following advantages.
(i) In the IPT+parquet method, the dynamical effects of the two-particle bosonic fluctuation in the full vertex, 
which give the diagonal structure and are more important in the multi-band systems,
are taken into account, 
whereas these are not  treated in the IPT. 
(ii) It is easy to apply diagrammatic extensions of DMFT for the non-local correlation. 
In the IPT+parquet method,
since the two-particle fluctuations in each channel (ph,${\rm \overline{ph}}$,pp) in the full vertex are estimated with a physically reasonable method (the parquet formalism), 
we can calculate the two-particle quantities necessary for diagrammatic extensions from this full vertex~[we show this point in a separate publication. Ref.~\cite{mizuno_s2fedf}]. 
On the contrary, in IPT, 
since there is no perspective on the two-particle fluctuations in the full vertex, 
it is difficult to estimate the two-particle quantities. 
Although a method to reconstruct the full vertex in IPT has been suggested in single-band systems~\cite{doi:10.1143/JPSJ.75.054713}, 
it cannot be used in multi-band systems.

IPT+parquet has a great advantage over CT-QMC in terms of the computational efficiency. 
We compare the cost of IPT+parquet with that of CT-QMC, employing the ``core hours''
\footnote{
  (core hours) = (the number of CPU cores we use) $\times$ (the number of hours for a calculation)
}
as an indicator of numerical costs. 
The order of core hours of IPT+parquet is ${\cal O}(1)$ and CT-QMC ${\cal O}(10^{2})$ 
in two-band cases.
The difference between the costs of the two methods may increase when we apply these to a system which has the larger degrees of freedom or is more realistic.
The CT-QMC simulations can suffer from the sign problem in these systems and need more and more samplings to obtain  reliable results while IPT+parquet does not have these difficulties.

\subsection{Scopes of applications of  MO-IPT and IPT+parquet}
Here,
we discuss the scopes of application of MO-IPT and IPT+parquet from the results of this study and the previous MO-IPT benchmark~\cite{Dasari2016}.

First,
we discuss the orbital degenerate systems with the interaction having only the intra-orbital elements.
The covalent insulator model in Sec.~3.3 in Ref.~\cite{Dasari2016}  
and
the bilayer model in Sec.~\ref{sec:2020-12-21-16-02} in this study 
correspond to these systems.
In these systems,
both MO-IPT and IPT+parquet show good agreements with the numerically exact CT-QMC.
As shown in Sec.~\ref{sec:2020-12-21-16-02},
IPT+parquet~(MO-IPT) is better near~(away from) half-filling.

Next,
we discuss the orbital degenerate systems with both intra- and inter-orbital interactions. 
The two-orbital Bethe lattice model in Sec.~3.4.1 and 3.4.2 of Ref.~\cite{Dasari2016} 
and 
the two-orbital square lattice model in Sec.~\ref{sec:2020-12-20-13-04} of our study
correspond to these systems.
As shown in Ref.~\cite{Dasari2016} and also in Sec.~\ref{sec:2020-12-20-13-04} of our study,
the MO-IPT results significantly deviate from that of the CT-QMC results at or near half filling.
This may be due to the drawback of the approximation in the modified parameter $B_{\alpha}$ in Eq.~(\ref{eq:2020-06-11-00-16}).
The calculation results lose electron-hole symmetry even in situations where the symmetry should be present
since $B=0$, 
which is the condition required in electron-hole symmetric systems [see Appendix~\ref{sec:2021-01-06-00-42} for details],
is not satisfied. 
Another version of MO-IPT~\cite{PhysRevLett.91.156402},
which was developed by Laad \textit{et al.} and has been applied to realistic systems~\cite{PhysRevB.90.115146,Koley2017,PhysRevB.100.121101,PhysRevB.73.045109,PhysRevB.96.165412,PhysRevB.100.115156},
does not have this drawback.
Hence,
systematic benchmarks of this version of MO-IPT for simple models,
which have not been performed to our knowledge, 
are desired.
If this version of MO-IPT 
turns out to also give results that deviate from CT-QMC at or near half-filling,
it may imply the limitations of the correction to modified parameters $A,B$ by the static many-particle correlation functions.
On the other hand, 
the IPT+parquet,
in which the many-particle correlation effects are considered as dynamical functions obtained by the parquet equations,
agrees well with CT-QMC at or near half-filling, 
as shown for the orbital-degenerate cases of the two-orbital square lattice and two-orbital Bethe lattice models in Sec.~\ref{sec:2020-12-20-13-04} and Sec.~\ref{sec:2021-01-06-01-35}, respectively.

Finally,
we discuss the orbital non-degenerate systems
which correspond to 
the two-orbital square lattice model in Sec.~\ref{sec:2020-12-20-13-04}
\footnote{  
  Dasari \textit{et al.} also studied an orbital non-degenerate system in Ref.~\cite{Dasari2016}, 
  where MO-IPT shows good agreement with CT-QMC for cases away from half-filling.
}
.
As mentioned in Sec.~\ref{sec:2020-12-20-13-04},
a remarkable feature in this situation is that 
the deviations of MO-IPT from CT-QMC are largely different between the two orbitals.
Namely, 
MO-IPT fails to appropriately  describe the correlation effects near half-filling that can depend on the orbitals when the orbitals are non-equivalent.
This can be understood from the new interpretation introduced in Sec.~\ref{sec:2020-12-25-04-04}.
As explained in Sec.~\ref{sec:2021-06-24-14-06}, 
$n_{0\alpha}=n_{\alpha}$ needs to be satisfied to estimate the correlation effects appropriately in each orbital.
However,
in MO-IPT,
$n_{0\alpha}$ and $n_{\alpha}$ do not satisfy this condition.
We overcome this difficulty by adding a degree of freedom to the pseudo chemical potential $\mu_{0}$ in IPT+parquet.
As a result,
the IPT+parquet agrees well with CT-QMC also in  non-degenerate systems.

\section{Conclusion}\label{sec:2020-12-25-04--13}
We have re-interpreted IPT 
as an approximation which captures the strong correlation effects by the mimicking the cross and central structures of the exact full vertex
and 
extended it such that it can be applied to the multi-band systems. 
We have validated this method~(IPT+parquet) by comparing it with the numerically exact CT-QMC method.
As a result,
we have confirmed that the results of IPT+parquet show good agreement with that of CT-QMC 
not only in the single-band systems but also in the multi-band systems.
In addition, numerical costs are largely reduced: core hours of IPT+parquet is at least 100 times smaller than that of CT-QMC.
We expect that IPT+parquet can be useful for analyzing various multi-band and strongly correlated systems.

\begin{acknowledgements}
  Part of the numerical calculations was performed using the large-scale computer systems provided by the following institutions: 
  the supercomputer center of the Institute for Solid State Physics, the University of Tokyo, 
  and 
  the Information Technology Center, the University of Tokyo.
  This study has been supported by JSPS KAKENHI Grants No.JP18H01860.
\end{acknowledgements}


\appendix

\section{Outline of DMFT and IPT solver}\label{sec:2020-12-25-04-00}

\subsection{DMFT}
DMFT is based on the equivalency between two models which is exact in the limit of the infinite spatial dimension $d \to \infty$.
In the finite spatial dimension case, 
DMFT can be considered as the approximation in which the temporal fluctuation is treated correctly instead of ignoring the spatial fluctuation.

In DMFT,
the lattice problem is solved by mapping it onto an impurity problem.
The lattice model~(Hubbard model) is given in Eq.~(\ref{eq:2020-05-12-23-57}).
The Anderson impurity model for multi-band systems is described as 
\begin{align}
  H 
  =&  \sum_{\bm{k}}\sum_{\alpha\beta} \epsilon_{\bm{k}\alpha\beta}b^{\dagger}_{\bm{k}\alpha}b_{\bm{k}\beta}
  +
  \sum_{\bm{k}}\sum_{\alpha\beta} ( V_{\bm{k}\alpha\beta}b^{\dagger}_{\bm{k}\alpha}f_{\beta} + {\rm h.c.} ) \nonumber \\
  &
  +
  \sum_{\alpha\beta}\epsilon_{f\alpha\beta}f^{\dagger}_{\alpha}f_{\beta}
  +
  \dfrac{1}{4}\sum_{\alpha\beta\gamma\lambda}U_{\alpha\beta\gamma\lambda} f^{\dagger}_{\alpha}f^{\dagger}_{\lambda}f_{\gamma}f_{\beta},
  \label{eq:2020-05-28-03-08}
\end{align}
where 
$b_{\bm{k}\alpha}^{(\dagger)}$ is annihilation~(creation) operator for bath electrons 
and 
$f^{(\dagger)}_{\alpha}$ for impurity electrons.
$\epsilon_{\bm{k}\alpha\beta}$ and $\epsilon_{f\alpha\beta}$ are the energy of bath and impurity electrons, respectively.
$V_{\bm{k}\alpha\beta}$ is the hybridization strength of bath and impurity electrons
and 
$U_{\alpha\beta\gamma\lambda}$ is the interaction in the impurity site.
The Green's function in the Hubbard model can be written as 
\begin{align}
  \hat{G}_{\rm lat}(k) 
  =&
  \bigl[ (i\omega_{n} + \mu)\hat{I} - \hat{\epsilon}_{\bm{k}} - \hat{\Sigma}(i\omega_{n}) \bigr]^{-1},
  \label{eq:2020-09-29-17-34}
\end{align}
where 
the spatial fluctuation~($\bm{k}$ dependence of the self energy) is ignored.
Also, 
the Green's function in the impurity model can be written as 
\begin{align}
  \hat{G}_{\rm imp}(i\omega_{n})
  =&
  \bigl[ (i\omega_{n} + \mu)\hat{I} - \hat{\epsilon}_{f} - \hat{\Delta}(i\omega_{n}) - \hat{\Sigma}(i\omega_{n}) \bigr]^{-1},
  \label{eq:2020-09-29-17-35}
\end{align}
where 
$\Delta(i\omega_{n})=N_{\bm{k}}^{-1}\sum_{\bm{k}}V_{\bm{k}}(i\omega_{n}-\epsilon_{\bm{k}})^{-1}V_{\bm{k}}$ is the hybridization function.
The self consistent condition in DMFT is given by 
\begin{align}
  \hat{\Delta}(i\omega_{n}) 
  =&
  (i\omega_{n} + \mu)\hat{I} - \hat{\Sigma}(i\omega_{n}) - \Bigl( \sum_{\bm{k}}\hat{G}_{\rm lat}(k) \Bigr)^{-1}.
  \label{eq:2020-09-29-17-55}
\end{align}
Although there are small differences depending on the impurity solvers, 
the actual calculation procedure in DMFT is roughly as follows.
\begin{enumerate}
  \item
    Start from an initial guess for the self energy $\Sigma(i\omega_{n})$.
  \item
    Calculate the lattice Green's function $G_{\rm lat}(k)$ by Eq.~(\ref{eq:2020-09-29-17-35}).
  \item
    Calculate the hybridization function $\Delta(i\omega_{n})$ by Eq.~(\ref{eq:2020-09-29-17-55}).
  \item
    Solve the impurity problem using $\Delta(i\omega_{n})$ and obtain the new self energy.
  \item
    Go back to step 2. (iterate until convergence).
\end{enumerate}
Various methods to solve the impurity problem (step 4) have been proposed~\cite{doi:10.1143/PTPS.46.244, doi:10.1143/PTP.53.970, doi:10.1143/PTP.53.1286, Yamada4, PhysRevB.45.6479,PhysRevLett.77.131,PhysRevB.55.16132,PhysRevB.86.085133,Saso_2001,doi:10.1143/JPSJ.72.777,PhysRevLett.91.156402,Dasari2016,PhysRevLett.97.076405,PhysRevB.76.235123,PhysRevB.74.155107,PhysRevB.72.035122,Rubtsov2004,Kuramoto1983,RevModPhys.80.395,PhysRevLett.72.1545,PhysRevB.86.165128}
and they are called impurity solvers.
We introduce two impurity solvers we use in this study in the following section.

\subsection{Iterative Perturbation Theory (IPT)}\label{sec:2021-01-01-01-12}
In the iterative perturbation theory (IPT)~\cite{doi:10.1143/PTPS.46.244, doi:10.1143/PTP.53.970, doi:10.1143/PTP.53.1286, Yamada4, PhysRevB.45.6479},
the correlation part of the self energy is approximated by the second order perturbation,
i.e.
\begin{align}
  \Sigma(i\omega_{n}) =& \Sigma^{\rm HF} + {\Sigma}^{\rm CR}(i\omega_{n}),
  \label{eq:2020-04-28-23-13} \\
  {\Sigma}^{\rm CR}(i\omega_{n}) \approx& \Sigma^{\rm 2nd}(i\omega_{n}) = T\sum_{\nu_{m}} U\chi_{0}(i\nu_{m})UG_{0}(i\omega_{n}+i\nu_{m}),
  \label{eq:2020-04-28-23-14} \\
  \chi_{0}(i\nu_{m}) =& -T\sum_{n}G_{0}(i\omega_{n})G_{0}(i\omega_{n}+i\nu_{m}),
  \label{eq:2020-04-28-23-15}\\
  G_{0}(i\omega_{n}) =& [ i\omega_{n} + \mu_{0} - \Delta(i\omega_{n}) - \Sigma^{\rm HF} ]^{-1},
  \label{eq:2020-04-28-23-16}
\end{align}
where
$\Sigma^{\rm HF}$ is the Hartree-Fock term (mean field term) 
and 
$\Sigma^{\rm CR}(i\omega_{n})$ the correlation term of the self energy, 
$\Delta(i\omega)$ the hybridization function,
$\mu_{0}$ the pseudo chemical potential.
In the electron-hole symmetric case, 
IPT provides a good result in both weak and strong correlation regimes.
Especially in the strong correlation limit, the IPT self energy reproduces the exact solution even though it is a perturbation solution from the weak coupling limit.
In other cases, 
however,
the results are not so good.
To overcome this weakness, 
modified-IPT~\cite{PhysRevLett.77.131,PhysRevB.55.16132,PhysRevB.86.085133} was proposed as an extended version of IPT for arbitrary filling.
In the modified-IPT,
the correlation part of the self energy is parametrized by 
\begin{align}
  {\Sigma}^{\rm CR}(i\omega_{n}) =& \dfrac{A \Sigma^{\rm 2nd}(i\omega_{n})}{1 - B\Sigma^{\rm 2nd}(i\omega_{n})}.
\end{align}
The constants $A$ and $B$ are determined such that one reproduces the exact solutions in the high frequency and atomic limits: 
\begin{align}
  A =& \dfrac{n(1-n)}{n_{0}(1-n_{0})}, \hspace{10pt} B = \dfrac{(1-2n)U + \mu_{0} - \mu}{n_{0}(1-n_{0})U^{2}},
  \label{eq:2020-05-31-16-54}
\end{align}
where 
$n_{0}$ and $n$ are the electron numbers evaluated from $G_{0}(i\omega_{n})$ and $G(i\omega_{n})$, respectively. 
The chemical potential $\mu$ is determined by fixing $n$ at the input value, 
while 
the pseudo chemical potential $\mu_{0}$ is a free parameter.
Various conditions for determining $\mu_{0}$ have been suggested: Luttinger sum rule, $n=n_{0}$ and so on~\cite{PhysRevLett.77.131,PhysRevB.55.16132,PhysRevB.86.085133}.
Hereafter, 
if we write IPT,
it refers to modified-IPT.

Furthermore, 
some extended versions of the modified-IPT for multi-orbital systems have been proposed~\cite{Saso_2001,doi:10.1143/JPSJ.72.777,PhysRevLett.91.156402,Dasari2016}.
Here, we summarize the outline of the MO-IPT developed in Ref.~\cite{Dasari2016}, 
which is the latest version of these methods,
with a slight modification.
In the MO-IPT, 
the orbital-diagonal parts of the self energy are parametrized as  
\begin{align}
  {\Sigma}^{\rm CR}_{\alpha\alpha}(i\omega_{n}) =& \dfrac{A_{\alpha} \Sigma^{\rm 2nd}_{\alpha\alpha}(i\omega_{n})}{1 - B_{\alpha}\Sigma^{\rm 2nd}_{\alpha\alpha}(i\omega_{n})},
  \label{eq:2020-04-29-18-44}
\end{align}
where $\alpha$ indicates the degrees of freedom of spin and orbital. 
Here, off-diagonal parts are ignored.  
Similarly to the single orbital case, 
$A_{\alpha}$ is determined such that one reproduces the exact solution in the high frequency limit.
On the other hand, 
$B_{\alpha}$ is determined such that one reproduces the approximate solution in the atomic limit 
since the exact solution can not be written in a simple form in multi-orbital systems.
Namely,
\begin{align}
  A_{\alpha}
  =&
  \dfrac{1}{\tau_{\alpha}}\sum_{\beta\neq\alpha}U_{\alpha\beta}\braket{n_{\beta}}(1-\braket{n_{\beta}})U_{\beta\alpha} \nonumber \\
  &+
  \dfrac{1}{\tau_{\alpha}}\sum_{\beta\neq\alpha}\sum_{\gamma\neq\beta\neq\alpha} U_{\alpha\beta} ( \braket{n_{\beta}n_{\gamma}} - \braket{n_{\beta}}\braket{n_{\gamma}} ) U_{\gamma\alpha}, 
  \label{eq:2020-06-11-00-10} \\
  B_{\alpha}
  =&
  \dfrac{1}{\tau_{\alpha}}\left( \mu_{0} -\mu-2\sum_{\beta(\neq\alpha)}U_{\alpha\beta}\braket{n_{\beta}} \right) \nonumber \\
  +&
  \dfrac{1}{\tau_{\alpha}^{2}A_{\alpha}}
  \sum_{\beta\gamma\eta(\neq \alpha)}U_{\alpha\beta}U_{\alpha\gamma}U_{\alpha\eta}
  \bigl( 
    \braket{n_{\beta}n_{\gamma}n_{\eta}}
    -
    \braket{n_{\beta}}\braket{n_{\gamma}n_{\eta}}
  \bigr),
  \label{eq:2020-06-11-00-16}\\
  \tau_{\alpha}
  =&
  \sum_{\beta}U_{\alpha\beta} \braket{n_{0\beta}}(1-\braket{n_{0\beta}}) U_{\beta\alpha},
  \label{eq:2020-06-11-00-17}
\end{align} 
where $U_{\alpha\beta} = U_{\alpha\alpha\beta\beta}$.
The difference between Eq.~(\ref{eq:2020-06-11-00-16}) and Eq.~(A.22) in Ref.~\cite{Dasari2016} is due to the difference between the notations of the zeroth-order Green's function.
Adopting the Matsubara frequency formalism,
we impose the condition $n_{\rm 0total}=n_{\rm total}$ to fix $\mu_{0}$, where $n_{\rm 0total}$ and $n_{\rm total}$ are the total electron density obtained from $G_{0}$ and $G$, 
while the real frequency is used in Ref.~\cite{Dasari2016} and $\mu_{0}$ is determined such that the Luttinger theorem is satisfied.
We have confirmed that the results are nearly independent of the adopted frequency types or conditions for $\mu_{0}$, 
by performing calculations for several models studied in Ref.~\cite{Dasari2016}.

\section{Diagrammatic origins of cross and central structures}\label{sec:2021-04-21-23-54}
\begin{figure}[]
  \centering
  \subfigure[A diagram independent of $i\omega_{n'}$. ]
  {\includegraphics[width=50mm,clip]{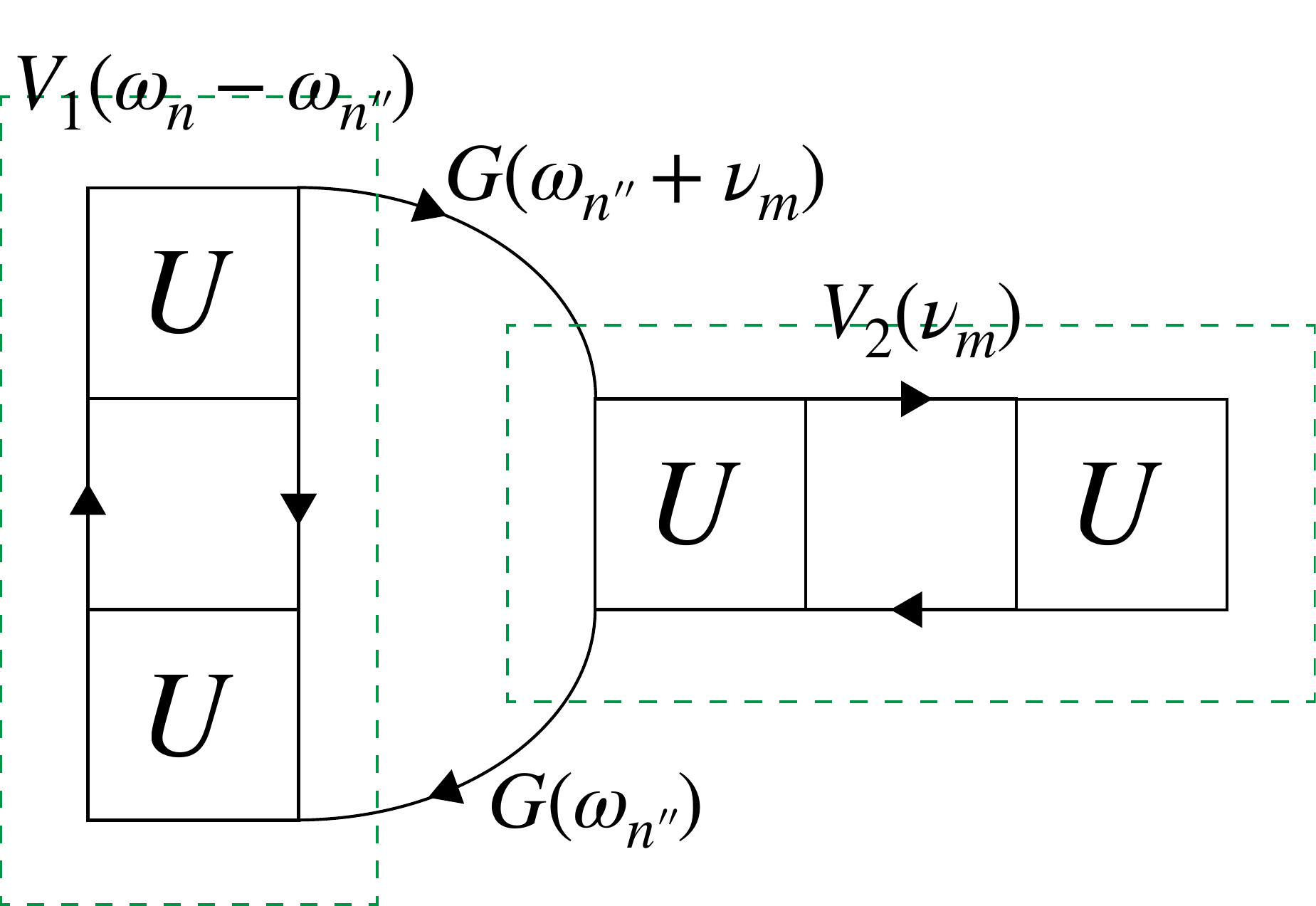}} \hspace{10pt}
  \subfigure[A diagram dependent on $i\omega_{n}$ and $i\omega_{n'}$ independently. ]
  {\includegraphics[width=70mm,clip]{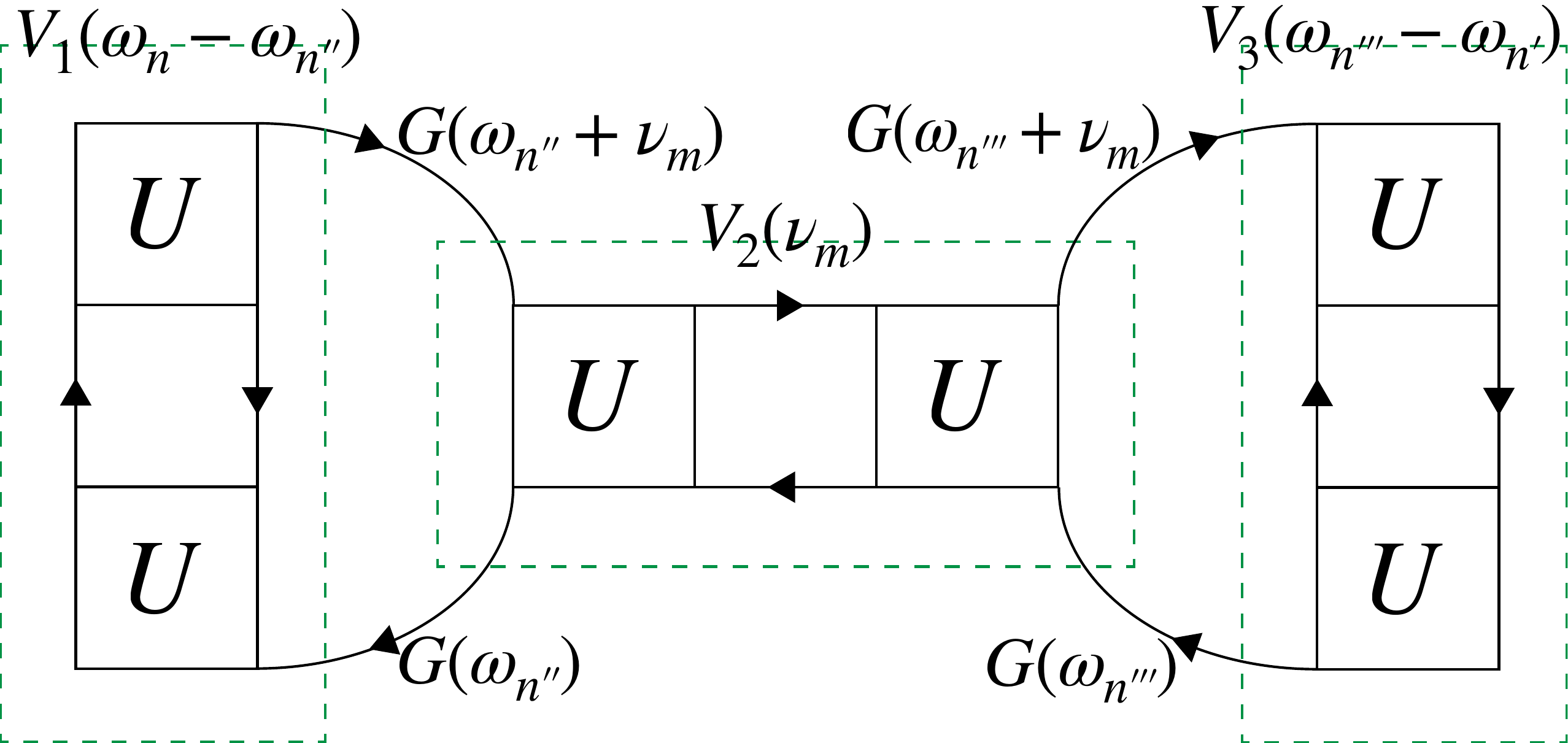}} 
  \caption{Examples of diagrams for each frequency structure.
  The diagrams depicted in (a) and (b) give the cross and  central structures, respectively. } 
  \label{fig:2020-06-14-13-48}
\end{figure}

Here, 
we show the diagrammatic origins of the cross and central structures.
First,
we explain the cross structure.
The combination of some channels depicted in Fig.~\ref{fig:2020-06-14-13-48}~(a), 
which extinguishes the $\omega_{n}$ or $\omega_{n'}$ dependence of $F^{\rm c}(i\omega_{n},i\omega_{n'},i\nu_{m})$ as follows. 
\begin{align}
  T \sum_{n''} V_{1}(\omega_{n}-\omega_{n''})G(\omega_{n''}+\nu_{m})G(\omega_{n''})V_{2}(\nu_{m})
  \label{eq:2020-06-27-16-10}
\end{align}
This contribution make $F^{\rm c}(i\omega_{n},i\omega_{n'},i\nu_{m})$
take large values in the vicinity of $\omega_{n}=0$ and $\omega_{n'}=0$ lines. 
Next,
we explain the central structure.
The multiple combinations of some channels depicted in Fig.~\ref{fig:2020-06-14-13-48}~(b)  yields the contribution which depends on $\omega_{n}$ and  $\omega_{n'}$ independently as
\begin{align}
  T^{2}\sum_{n'',n'''}
  &V_{1}(\omega_{n}-\omega_{n''})G(\omega_{n''}+\nu_{m})G(\omega_{n''})
  \nonumber \\
  &\times V_{2}(\nu_{m})
  G(\omega_{n'''}+\nu_{m})G(\omega_{n'''})
  V_{3}(\omega_{n'''}-\omega_{n'}).
  \label{eq:2020-06-27-16-11}
\end{align}
This makes $F^{\rm c}(i\omega_{n},i\omega_{n'},i\nu_{m})$
take large values in the center of $n-n'$ plane. 

As we can see from its origin,
the cross or central structures come from the higher order diagrams than that of the diagonal structure.
Therefore these contributions are important in the strongly correlated regime.

Also, 
we can show that the simple product form of the full vertex in Eqs.~(\ref{eq:2020-08-28-01-28}) and (\ref{eq:2020-08-28-01-09}) is reasonable to some extent
by considering the nature of the two-particle functions which transfer the bosonic frequencies $\nu_{m}$.
For example, $V_{i}(\nu_{m})$ in Eq.~(\ref{eq:2020-06-27-16-10}) or Eq.~(\ref{eq:2020-06-27-16-11}) has a large value in the vicinity of $\nu_{m}=0$
and becomes similar to the $\delta$ function when the two-particle fluctuation becomes large.
As an extreme case, if we approximate $V_{1}$ and $V_{3}$ by the $\delta$ function in Eq.~(\ref{eq:2020-06-27-16-10}) or Eq.~(\ref{eq:2020-06-27-16-11}), 
we obtain the frequency dependence of ph part in Eqs.~(\ref{eq:2020-08-28-01-28}) and (\ref{eq:2020-08-28-01-09}).

\section{Full vertex in the atomic limit}\label{sec:2021-04-03-00-07}
In the case of single-band systems,
we can write down the full vertex in the atomic limit as follows~\cite{PhysRevB.86.125114}.
\begin{align}
  F^{c,s}_{\rm atom} =& F^{\uparrow\uparrow}_{\rm atom} \pm  F^{\uparrow\downarrow}_{\rm atom} \label{eq:2020-06-04-11-25} \\
  F^{\uparrow\uparrow}_{\rm atom} 
  =&
  -\beta \dfrac{U^{2}}{4} (\delta_{\omega_{n}\omega_{n'}}-\delta_{\nu_{m} 0} ) \nonumber \\
  &-\beta \dfrac{U^{4}}{16} (\delta_{\omega_{n}\omega_{n'}}-\delta_{\nu_{m} 0} ) \Bigl( \dfrac{1}{\omega_{n}^{2}} + \dfrac{1}{(\omega_{n'}+\nu_{m})^{2}} \Bigr) \nonumber \\
  &-\beta \dfrac{U^{6}}{64}\dfrac{\delta_{\omega_{n}\omega_{n'}}-\delta_{\nu_{m} 0} }{\omega_{n}^{2}(\omega_{n'}+\nu_{m})^{2}} 
  \label{eq:2020-06-04-11-28} \\
  F^{\uparrow\downarrow}_{\rm atom} 
  =&
  U
  + \beta \dfrac{U^{2}}{4} \Bigl[  
    \dfrac{2\delta_{\omega_{n}(-\omega_{n'}-\nu_{m})}+\delta_{\nu_{m} 0}}{1+e^{\beta U/2}}
    -
    \dfrac{2\delta_{\omega_{n}\omega_{n'}}+\delta_{\nu_{m} 0}}{1+e^{-\beta U/2}}
  \Bigr]
  \nonumber \\
  &+ \dfrac{U^{3}}{8} \dfrac{\omega_{n}^{2}+(\omega_{n}+\nu_{m})^{2}+(\omega_{n'}+\nu_{m})^{2}+\omega_{n'}^{2}}{\omega_{n}(\omega_{n}+\nu_{m})(\omega_{n'}+\nu_{m})\omega_{n'}}
  \nonumber \\
  &+ \beta \dfrac{U^{4}}{16} \Bigl[  
    \dfrac{2\delta_{\omega_{n}(-\omega_{n'}-\nu_{m})}+\delta_{\nu_{m} 0}}{1+e^{\beta U/2}}
    \Bigl( \dfrac{1}{(\omega_{n}+\nu_{m})^{2}} + \dfrac{1}{(\omega_{n'}+\nu_{m})^{2}} \Bigr)
    \nonumber \\
    &-
    \dfrac{2\delta_{\omega_{n}\omega_{n'}}+\delta_{\nu_{m} 0}}{1+e^{-\beta U/2}}
    \Bigl( \dfrac{1}{\omega_{n}^{2}} + \dfrac{1}{(\omega_{n'}+\nu_{m})^{2}} \Bigr)
  \Bigr]
  \nonumber \\ 
  & + \dfrac{3U^{5}}{16}\dfrac{1}{\omega_{n}(\omega_{n}+\nu_{m})(\omega_{n'}+\nu_{m})\omega_{n'}}
  \nonumber \\
  &+ \beta \dfrac{U^{6}}{64} \Bigl[  
    \dfrac{2\delta_{\omega_{n}(-\omega_{n'}-\nu_{m})}+\delta_{\nu_{m} 0}}{1+e^{\beta U/2}}
    \dfrac{1}{(\omega_{n}+\nu_{m})^{2}(\omega_{n'}+\nu_{m})^{2}} 
    \nonumber \\
    &-
    \dfrac{2\delta_{\omega_{n}\omega_{n'}}+\delta_{\nu_{m} 0}}{1+e^{-\beta U/2}}
    \dfrac{1}{\omega_{n}^{2}(\omega_{n'}+\nu_{m})^{2}}
  \Bigr]
  \label{eq:2020-06-04-11-29} 
\end{align}
The structure of each order is as follows.
\begin{itemize}
  \item $U^{1} \to $ constant 
  \item $U^{2} \to $ diagonal structure
  \item $U^{3} \to $ cross structure 
  \item $U^{4} \to $ diagonal $\times$ cross structure
  \item $U^{5},U^{6} \to $ diagonal $\times$ central structure
\end{itemize}
We can indeed see that the diagonal structure is dominant in the small $U$ regime and the cross and central structures develop as $U$ increases. 

\section{Simplified parquet method}\label{sec:2020-10-09-00-10}
In this section,
we introduce the simplified parquet method developed in Ref.~\cite{doi:10.1143/JPSJ.79.094707}, 
in which the numerical cost is much lower than that of the non-simplified parquet method since 
we should practically consider just one of the three variables $(k,k',q)$.
Since the simplified parquet method in Ref.~\cite{doi:10.1143/JPSJ.79.094707}
has not been extended for multi-band systems,
we extend it for our purpose.
We show here this multi-band version of the simplified parquet method.
There are differences between coefficients in Ref.~\cite{doi:10.1143/JPSJ.79.094707} and in our notation.
These come from the difference in treatments of $1/2$ factor which is needed to avoid the double counting of diagrams in pp channel.  
This factor emerges in the definitions of the vertices in our notation while it emerges in the Bethe-Salpeter equation in Ref.~\cite{doi:10.1143/JPSJ.79.094707}.
Before we start introducing the simplified parquet method,
we define the following notation which indicates the set of the degrees of freedom, the frequencies, and wave vectors.
\begin{align}
  D =& (\alpha,\beta,\gamma,\lambda), (k,k',q) \label{eq:2020-05-10-14-49} \\
  C =& (\alpha,\gamma,\beta,\lambda), (k,k+q,k'-k) \label{eq:2020-05-10-14-51} \\
  P =& (\alpha,\lambda,\gamma,\beta), (k,k',-q-k-k') \label{eq:2020-05-10-14-52} \\
  X =& (\alpha,\gamma,\lambda,\beta), (k,-k-q,k'-k) \label{eq:2020-05-10-14-53}
\end{align}

In the presence of SU(2) symmetry in spin space, 
the full vertex can be divided into four channels 
c(charge), s(spin), e(even), o(odd)
in terms of the parity of spin.
\begin{align}
  F_{r}(D) =& \Lambda_{r}(D) + \Phi_{ {\rm ph}, r}(D) + \Phi_{ {\rm \overline{ph}},r}(D) + \Phi_{ {\rm pp},r}(D) \hspace{10pt} (r = {\rm c,s,e,o} )
  \label{eq:2020-06-14-15-03}
\end{align} 
We can rewrite 
Eq.~(\ref{eq:2020-06-14-15-03})
as follows by replacements of variables and indices.
\begin{align}
  F_{\rm c}(D) =& \Lambda_{\rm c}(D) + \Phi_{\rm ph,c}(D) \nonumber \\ &\hspace{-20pt}- \dfrac{1}{2}[ \Phi_{\rm ph,c} + 3\Phi_{\rm ph,s} ](C)  + [ \Phi_{\rm pp,e} - 3\Phi_{\rm pp, o} ](P) \label{eq:2020-05-12-13-58} \\  
  F_{\rm s}(D) =& \Lambda_{\rm s}(D) + \Phi_{\rm ph,s}(D) \nonumber \\ &\hspace{-20pt}- \dfrac{1}{2}[ \Phi_{\rm ph,c} -  \Phi_{\rm ph,s} ](C)  - [ \Phi_{\rm pp,e} -  \Phi_{\rm pp, o} ](P) \label{eq:2020-05-12-13-59} \\  
  F_{\rm e}(D) =& \Lambda_{\rm e}(D) + \Phi_{\rm pp,e}(D) \nonumber \\ &\hspace{-20pt}+ \dfrac{1}{4}[ \Phi_{\rm ph,c} - 3\Phi_{\rm ph,s} ](X)  + \dfrac{1}{4}[ \Phi_{\rm ph,c} - 3\Phi_{\rm ph, s} ](P) \label{eq:2020-05-12-14-00} \\  
  F_{\rm o}(D) =& \Lambda_{\rm o}(D) + \Phi_{\rm pp,o}(D) \nonumber \\ &\hspace{-20pt}+ \dfrac{1}{4}[ \Phi_{\rm ph,c} +  \Phi_{\rm ph,s} ](X)  - \dfrac{1}{4}[ \Phi_{\rm ph,c} + \Phi_{\rm ph, s} ](P) \label{eq:2020-05-12-14-01}   
\end{align} 
As we can see from 
Eqs.~(\ref{eq:2020-05-12-13-58})$-$(\ref{eq:2020-05-12-14-01}),
since c,s is always together with ph, 
e,o with pp, 
we omit the subscripts ph or pp hereafter. 
And we write the third and fourth term as $\gamma^{(1)}_{r}$ and $\gamma^{(2)}_{r}$, respectively.
To say, 
\begin{align}
  \gamma^{(1)}_{\rm c}(D) =& - \dfrac{1}{2}[ \Phi_{\rm c} + 3\Phi_{\rm s} ](D) \label{eq:2021-04-27-01-02} \\ 
  \gamma^{(1)}_{\rm s}(D) =& - \dfrac{1}{2}[ \Phi_{\rm c} -  \Phi_{\rm s} ](D) \label{eq:2021-04-27-01-03} \\
  \gamma^{(1)}_{\rm e}(D) =& \dfrac{1}{4}[ \Phi_{\rm c} - 3\Phi_{\rm s} ](D) \label{eq:2021-04-27-01-04} \\
  \gamma^{(1)}_{\rm o}(D) =& \dfrac{1}{4}[ \Phi_{\rm c} +  \Phi_{\rm s} ](D) \label{eq:2021-04-27-01-05} 
\end{align}
\begin{align}
  \gamma^{(2)}_{\rm c}(D) =&  [ \Phi_{\rm e} - 3\Phi_{\rm  o} ](D) \label{eq:2021-04-27-01-06} \\
  \gamma^{(2)}_{\rm s}(D) =&  - [ \Phi_{\rm e} -  \Phi_{\rm  o} ](D)\label{eq:2021-04-27-01-07} \\
  \gamma^{(2)}_{\rm e}(D) =&   \dfrac{1}{4}[ \Phi_{\rm c} - 3\Phi_{\rm s} ](D)\label{eq:2021-04-27-01-08}\\
  \gamma^{(2)}_{\rm o}(D) =&  - \dfrac{1}{4}[ \Phi_{\rm c} + \Phi_{\rm s} ](D)\label{eq:2021-04-27-01-09}
\end{align}

We can also write the 
Bethe-Salpeter equation by using the four channels: 
\begin{align}
  \hat{F}_{r} &= \hat{\Gamma}_{r} + \hat{\Phi}_{r} \hspace{10pt} (r = {\rm c,s,e,o}) \label{eq:2020-05-10-13-33} \\
  \hat{\Phi}_{r} &= -\hat{\Gamma}_{r}\hat{\chi}_{0} \hat{F}_{r} = -\hat{\Gamma}_{r}\hat{\chi}_{r}\hat{\Gamma}_{r},
  \label{eq:2020-05-10-13-34} 
\end{align}
and the susceptibilities: 
\begin{align} 
  \hat{\chi}_{r} =& \hat{\chi}_{0} - \hat{\chi}_{0}\hat{\Gamma}_{r}\hat{\chi}_{r} = \hat{\chi}_{0} - \hat{\chi}_{0} \hat{F}_{r}\hat{\chi}_{0}.
  \label{eq:2020-05-12-14-57}
\end{align}

With this preliminary, we will explain the details of the approximation in the simplified parquet method.
First, 
we use the bare vertices $U_{r}$ as the fully irreducible vertices $\Lambda_{r}$:
\begin{align}
  \Lambda_{\rm c}(D) =& (U_{\sigma\sigma\sigma\sigma} + U_{\sigma\sigma\bar{\sigma}\bar{\sigma}})(D) = U_{\rm c}(D) \label{eq:2020-05-12-15-37} \\
  \Lambda_{\rm s}(D) =& (U_{\sigma\sigma\sigma\sigma} - U_{\sigma\sigma\bar{\sigma}\bar{\sigma}})(D) = -U_{\rm s}(D) \label{eq:2020-05-12-15-38} \\
  \Lambda_{\rm e}(D) =& \dfrac{1}{2}(U^{\rm pp}_{\sigma\bar{\sigma}\sigma\bar{\sigma}} - U^{\rm pp}_{\sigma\bar{\sigma}\bar{\sigma}\sigma})(D) = \dfrac{1}{4}(U_{\rm c} + 3U_{\rm s} )(P) \label{eq:2020-05-12-15-39} \\
  \Lambda_{\rm o}(D) =& \dfrac{1}{2}(U^{\rm pp}_{\sigma\bar{\sigma}\sigma\bar{\sigma}} + U^{\rm pp}_{\sigma\bar{\sigma}\bar{\sigma}\sigma})(D) = -\dfrac{1}{4}(U_{\rm c} - U_{\rm s} )(P) \label{eq:2020-05-12-15-40}
\end{align}
We calculate susceptibilities by using the random phase approximation~(RPA) type formula:
\begin{align} 
  \hat{\chi}_{r}(q) =& \hat{\chi}_{0}(q)[\hat{I} + \hat{\tilde{\Lambda}}_{r}\hat{\chi}_{0}(q)]^{-1}.
  \label{eq:2020-05-10-21-27} 
\end{align}
where 
\begin {align}
  \hat{\tilde{\Lambda}}_{r} =& z_{r} \hat{\Lambda}_{r} 
  \label{eq:2020-05-10-21-21}
\end{align}
and $z_{r}$ is the constant renormalization factor.
With these, 
the irreducible vertices can be calculated as 
\begin{align} 
  \hat{\Phi}_{r} =&
  -\hat{\tilde{\Lambda}}_{r} \hat{\chi}_{r} \hat{\tilde{\Lambda}}_{r} .
  \label{eq:2020-05-12-16-20} 
\end{align}
By this approximation,
the generalized momentum dependences in Eqs.~(\ref{eq:2020-05-10-14-49})-(\ref{eq:2020-05-10-14-53}) are replaced as
\begin{align}
  D :&\ (k,k',q) \to q \label{eq:2021-04-22-02-08} \\
  C :&\ (k,k+q,k'-k) \to k'-k \label{eq:2021-04-22-02-09} \\
  P :&\ (k,k',-q-k-k') \to -q-k-k'\label{eq:2021-04-22-02-10} \\
  X :&\ (k,-k-q,k'-k) \to k'-k \label{eq:2021-04-22-02-11}
\end{align}
If we consider the local case,
Eqs.~(\ref{eq:2021-04-22-02-08})-(\ref{eq:2021-04-22-02-11}) mean that the full vertex has only the diagonal structure.

From the comparison between
susceptibilities from the RPA type Eq.~(\ref{eq:2020-05-10-21-27}) and the parquet type Eq.~(\ref{eq:2020-05-12-14-57}),
we can obtain the renormalization factor $z_{r}$ as 
\begin{align}
  z_{r} =&
  1 + 
  \dfrac{ {\rm Tr} \bigl[ \hat{\chi}_{0}(k,q)( \hat{\gamma}^{(1)}_{r}(k-k') + \hat{\gamma}^{(2)}_{r}(k+k'+q) ) \hat{\chi}_{0}(k',q) \bigr]}{ {\rm Tr} \bigl[\hat{\chi}_{0}(q)\hat{\Lambda}_{r}\hat{\chi}_{0}(q)\bigr]},
  \label{eq:2020-05-10-16-29}
\end{align}
where ${\rm Tr}A = \sum_{k,k',q}\sum_{\alpha} A_{\alpha \alpha \alpha \alpha}(k,k',q)$. 
Although the summation in the numerator of Eq.~(\ref{eq:2020-05-10-16-29}) is taken over $k,k',q$, 
we can rewrite it as a summation over $q$ by a variable conversion.
Hence, 
we treat only $q$ practically.
The calculation procedure of the simplified parquet method is as follows.
\begin{enumerate}
  \item
    Calculate the bare vertices $\Lambda_{r}$  by Eqs.~(\ref{eq:2020-05-12-15-37})-(\ref{eq:2020-05-12-15-40}).
  \item
    Calculate the renormalized vertices $\tilde{\Lambda}_{r}$ by Eq.~(\ref{eq:2020-05-10-21-21}). \\ 
    The initial values are $(z_{\rm c},z_{\rm s},z_{\rm e},z_{\rm o})=(1,0.1,1,1)$.  
  \item
    Calculate the susceptibilities $\chi_{r}$ by Eq.~(\ref{eq:2020-05-10-21-27}).
  \item
    Calculate the reducible vertices $\Phi_{r}$ by Eq.~(\ref{eq:2020-05-12-16-20}).
  \item
    Calculate the vertices $\gamma^{(1)}_{r}$ and $\gamma^{(2)}_{r}$ by  Eqs.~(\ref{eq:2021-04-27-01-02})-(\ref{eq:2021-04-27-01-09}).
  \item
    Update the renormalization factor $z_{r}$ by Eq.(\ref{eq:2020-05-10-16-29}).
  \item
    Go back to step 2. (until convergence).
\end{enumerate}
After convergence, 
we already have obtained the vertices $\Phi_{r}$, $F_{r}$ and the susceptibilities $\chi_{r}$.

If we obtain the full vertex by the above procedure, we can obtain the self energy as follows.
\begin{align}
  \Sigma_{\alpha\beta}(k)
  =&
  \dfrac{1}{4}\sum_{\gamma\lambda}\sum_{q} \Bigl[  \hat{F}_{\rm c}(q) \hat{\chi}_{0}(q)\hat{U}_{\rm c} + 3\hat{F}_{\rm s}(q)\hat{\chi}_{0}(q)\hat{U}_{\rm s} \Bigr]_{\alpha\gamma\beta\lambda} G_{\gamma\lambda}(k+q) 
  \label{eq:2020-06-28-17-38}
\end{align}
In practical calculation, however, we omit the contribution from pp channel in self energy since it tends to be overestimated.

%

\

\section{The conditions of the modified parameters in the electron-hole symmetric case}\label{sec:2021-01-06-00-42}
Here, 
we show the conditions which the modified parameters $A,B$ need to satisfy in the electron-hole symmetric case.
For simplicity, we consider the single-orbital case.

The spectral representation of the Matsubara self energy is expressed as 
\begin{align}
  \Sigma(\omega_{n}) 
  =& 
  \dfrac{1}{\pi}\int_{-\infty}^{\infty}d\omega \dfrac{ (-{\rm Im}\Sigma(\omega))}{i\omega_{n}-\omega}.
  \label{eq:2021-01-05-19-50}
\end{align}
Hence,
the real part of the Matsubara self energy is 
\begin{align}
  {\rm Re}\Sigma(i\omega_{n})
  =&
  \dfrac{1}{\pi}\int_{-\infty}^{\infty}d\omega \dfrac{\omega}{\omega_{n}^{2}+\omega^{2}} {\rm Im}\Sigma(\omega).
  \label{eq:2021-01-05-19-54}
\end{align}
In the presence of the electron-hole symmetry,
${\rm Im}\Sigma(\omega)$ is an even function,
so ${\rm Re}\Sigma(i\omega_{n})=0$.

On the other hand, 
the ansatz of the modified-IPT self energy is 
\begin{align}
  \dfrac{A\Sigma^{\rm 2nd}(i\omega_{n})}{1 - B\Sigma^{\rm 2nd}(i\omega_{n})},
  \label{eq:2021-01-05-20-11}
\end{align}
where $\Sigma^{\rm 2nd}(i\omega_{n})$ is the second-order self energy.
In the presence of the electron-hole symmetry, 
the condition which $A$ and $B$ need to satisfy is 
\begin{align}
  A, iB \in \mathbb{R}
\end{align}
since $\Sigma^{\rm 2nd}(i\omega_{n})$ is a pure imaginary function.
When $ B$ is a real number,
$B$ needs to be zero.

\bibliographystyle{../../bibfiles/prb}
\bibliographystyle{apsrev4-1}
\bibliography{../../bibfiles/reference}


\end{document}